# Ubiquity of Fourier Transformation in Optical Sciences

Masud Mansuripur

College of Optical Sciences, The University of Arizona, Tucson, Arizona 85721

**Abstract**. The Fourier transform operation is an important conceptual as well as computational tool in the arsenal of every practitioner of physical and mathematical sciences. We discuss some of its applications in optical science and engineering, with the goal of providing a broad perspective on the intimate relation between the physical and mathematical concepts that are elegantly interwoven within the theory of Fourier transforms.

**1. Introduction**. Readers of *Applied Optics* are no strangers to the many applications of the Fourier transform theory in optical science and engineering, so, in a sense, this paper needs no introduction. The goal is to look at a few simple, yet profound, examples across a number of sub-disciplines of optical sciences, and to share with the reader the joy of learning and appreciating widely different subjects from the perspective offered by a unified underlying framework.

In the next section, the basic notion of Fourier transformation[1] is introduced and a simple proof of its fundamental theorem is given. Section 3 shows how Fourier transformation in two-dimensional space is used to describe the propagation of coherent light within the scalar theory of diffraction. Here, we use a deceptively simple argument based on the method of stationary-phase approximation[2] to arrive at the formula for the light amplitude profile in the far field (or Fraunhofer) diffraction regime.[3,4] The same formula is then shown to describe the focusing of coherent light through a conventional lens.

Section 4 examines the eigenfunctions and eigenvalues of the Fourier operator, showing that a large class of functions exists in which every function is its own Fourier transform; this, despite the fact that the Fourier operator admits only four possible eigenvalues: $\pm 1$ and $\pm$i.

Applications of the Fourier transform in classical electrodynamics are the subject of the next four sections. In Sec.5 we show that electromagnetic (EM) fields and their sources can be transformed from the $(\boldsymbol{r}, t)$ spacetime into the $(\boldsymbol{k}, \omega)$ Fourier domain using four-dimensional Fourier transformation. A review of Maxwell's equations in Sec.6 then reveals that the ever-present differential operators of the vector calculus, namely, gradient $(\nabla)$, divergence $(\nabla \cdot)$, curl $(\nabla \times)$, and Laplacian $(\nabla^2)$, altogether disappear in the Fourier domain, being replaced by ordinary vector algebraic operations. This enables a complete solution of Maxwell's equations in the Fourier domain, where the EM fields are straightforwardly related to their sources via simple algebraic equations. What remains then is a return to the $(\boldsymbol{r}, t)$ spacetime domain, which is accomplished by inverse Fourier transformation. The standard vector and scalar potentials, along with the important notion of gauge in electrodynamics, emerge rather naturally in Sec.7 from a consideration of Maxwell's equations.[5-11] Here, we use an elementary example to show the power of the Fourier domain techniques in explaining the generation of propagating as well as evanescent EM waves from a sheet of oscillating electrical currents. Several important aspects of EM radiation, such as the self-field and radiation resistance, the rates of radiation of energy and linear momentum, and the overall conservation of energy and momentum are explained through this unique example. Finally, in Sec.8, we show another application of the Fourier domain techniques by deriving general formulas for the Hertz vector potentials.[4]

Students of quantum mechanics and quantum optics are quite familiar with the numerous applications of Fourier transformation in analyzing quantum systems.[12,13] In fact, a most fundamental wave-function in Schrödinger's quantum mechanics, that of a point-particle of mass $m$, enery $\mathcal{E}$, and linear momentum $\boldsymbol{p}$ in free space, is $\exp[\mathrm{i}(\boldsymbol{k}\cdot\boldsymbol{r}-\omega t)]$, where $\boldsymbol{p}=\hbar\boldsymbol{k}$ and $\mathcal{E}=\hbar\omega$. This wave-function, of course, is nothing but the kernel of the Fourier operator in our four-dimensional spacetime. In Sec.9, we use the example of a point particle trapped in the quadratic potential well of a harmonic oscillator, to showcase a natural, beautiful, and unexpected appearance of Fourier transformation in a quantum system. Another interesting appearance of Fourier transformation in quantum mechanics is in the Wigner distribution, which, aside from quantum optics, has found recent applications in several other areas such as classical optics as well as electrical, optical, and acoustic signal processing.[14-18] Section 10 introduces the general idea of the Wigner function along with some of its basic properties and applications.

As an instance of the numerous applications of Fourier transformation in statistical optics, Sec.11 elaborates one of the fundamental results that has been discovered in this branch of optical science, namely, the van Cittert-Zernike theorem.[2,4] The theorem states that the degree of mutual coherence between two points in a plane illuminated by a distant, quasi-monochromatic, and spatially incoherent light source is given by the Fourier transform of the intensity distribution in the plane of the source.

Bandlimited functions are obviously important in practical applications, since any realistic function cannot be expected to contain infinitely large frequencies. An interesting feature of such functions, however, has come to light only in recent years and has found application in super resolution optical imaging and optical microscopy.[19-21] As it turns out, it is possible to construct bandlimited functions that contain arbitrarily long intervals over which they oscillate at frequencies well beyond their bandwidth—albeit with exceedingly weak amplitudes. While we will not discuss such superoscillatory functions here, we do describe an important characteristic of bandlimited functions in Sec.12, namely, the rate of decline as $|x|\to\infty$ of a function such as $f(x)$ whose Fourier transform $F(s)$ precisely vanishes outside its bandwidth, where $|s|>s_0$.

The paper closes with a few remarks and observations in Sec.13 pertaining to fractional differentiation using the Fourier transform. Although no practical applications of this concept are known to the author, he feels that the topic is interesting enough to deserve the reader's attention.

It goes without saying that many important applications of the Fourier transform theory in the fields of optics and photonics must, of necessity, be left out of the paper lest it becomes excessively long and unwieldy. Ignoring topics such as spectral pulse shaping, Fourier transform spectroscopy, optical stellar interferometry, far-field scattering, optical clocks, frequency combs, optical trapping and micromanipulation, the general theory of optical coherence, super resolution optical microscopy, etc., is by no means indicative of a lack of broad interest in these subjects. The goal of the paper, however, is not so much a comprehensive coverage as it is to showcase a few illustrative instances of Fourier analysis as applied in the optical sciences. Along the way, we aim to arrive at some well-known results either in a new way, or perhaps from a less commonly appreciated perspective.

**2. Preliminaries**. The Fourier transform $F(s)$ of a function $f(x)$, where $x$ and $s$ are real-valued variables, while $f(\cdot)$ and $F(\cdot)$ are, in general, complex-valued functions of their respective variables, is defined as follows:

$$F(s)=\int_{-\infty}^{\infty} f(x)\exp(-\mathrm{i}2\pi sx)\,\mathrm{d}x. \tag{1}$$

When the above integral exists, the function $f(x)$ is said to be Fourier transformable.[1] The inverse Fourier transform integral then reproduces the original function, as follows:

$$f(x) = \int_{-\infty}^{\infty} F(s) \exp(\mathrm{i}2\pi s x)\, \mathrm{d}s. \tag{2}$$

Occasionally, the exponential functions appearing in Eqs.(1) and (2) are interchanged, so that the forward transform has a $+$ sign in its exponent while the inverse transform has a $-$ sign.

---

**Example**. As an elementary example, let $f(x) = \exp(-\alpha|x| + \mathrm{i}2\pi s_0 x)$, where $s_0$ and $\alpha > 0$ are arbitrary real-valued constants. Equation (1) yields the Fourier transform of $f(x)$, as follows:

$$F(s) = \int_{-\infty}^{0} e^{[\alpha-\mathrm{i}2\pi(s-s_0)]x}\mathrm{d}x + \int_{0}^{\infty} e^{-[\alpha+\mathrm{i}2\pi(s-s_0)]x}\mathrm{d}x$$
$$= \frac{1}{\alpha - \mathrm{i}2\pi(s-s_0)} + \frac{1}{\alpha + \mathrm{i}2\pi(s-s_0)} = \frac{2\alpha}{4\pi^2(s-s_0)^2 + \alpha^2}. \tag{3}$$

Note that the area under the function $F(s)$, namely, $\int_{-\infty}^{\infty} F(s)\mathrm{d}s$, equals 1.0, independently of the value of $\alpha$. In the limit when $\alpha \to 0$, we have $F(s) \to \delta(s - s_0)$, revealing Dirac's delta-function $\delta(s - s_0)$ as the Fourier transform of $\exp(\mathrm{i}2\pi s_0 x)$. The inverse transform of $\delta(s - s_0)$ is readily seen from Eq.(2) via the sifting property of the $\delta$-function to be $\exp(\mathrm{i}2\pi s_0 x)$.

---

The identity $\delta(x - x') = \delta(x' - x) = \int_{-\infty}^{\infty} \exp[\pm\mathrm{i}2\pi(x - x')s]\, \mathrm{d}s$, which is an important corollary of the above example, may now be used to prove Fourier's main theorem, namely, the validity of the inverse Fourier integral in Eq.(2). Substituting Eq.(1) into Eq.(2), we find

$$f(x) = \int_{-\infty}^{\infty} \left[\int_{-\infty}^{\infty} f(x')e^{-\mathrm{i}2\pi s x'}\mathrm{d}x'\right] e^{\mathrm{i}2\pi s x}\mathrm{d}s = \int_{-\infty}^{\infty} f(x') \left[\int_{-\infty}^{\infty} e^{\mathrm{i}2\pi s(x-x')}\mathrm{d}s\right] \mathrm{d}x'$$
$$= \int_{-\infty}^{\infty} f(x')\delta(x' - x)\mathrm{d}x' = f(x). \tag{4}$$

Needless to say, this "proof" overlooks some important and subtle nuances of the Fourier transform theory, such as the conditions under which the integral in Eq.(1) exists, circumstances that allow the reversal of the order of integration in Eq.(4), the validity of the limiting argument that reduces the inner integral in Eq.(4) to a $\delta$-function, and the mathematical sense in which $f(x)$ on the left-hand side of Eq.(4) is said to be equal to $f(x)$ on the right-hand side.[1] Nevertheless, for functions of practical interest, the essential features of the proof hold and the inverse Fourier integral in Eq.(2) reconstitutes the original function $f(x)$.

**3. Fourier optics**. The two-dimensional version of Fourier transformation finds application in the field of Fourier Optics.[3,4,22] This is because the Helmholtz equation $(\nabla^2 - v^{-2}\partial_t^2)f(\boldsymbol{r}, t) = 0$, which governs the propagation of scalar waves in homogeneous media,[4] has eigen-solutions of the general form $f(\boldsymbol{r}, t) = a_0 \exp[\mathrm{i}(\boldsymbol{k} \cdot \boldsymbol{r} - \omega t)]$, where $a_0$ is an arbitrary complex constant, $\boldsymbol{k} = k_x\hat{\boldsymbol{x}} + k_y\hat{\boldsymbol{y}} + k_z\hat{\boldsymbol{z}}$ is an arbitrary complex vector, and $\boldsymbol{k} \cdot \boldsymbol{k} = k_x^2 + k_y^2 + k_z^2 = (\omega/v)^2$.[‡] Here, $\boldsymbol{r} = x\hat{\boldsymbol{x}} + y\hat{\boldsymbol{y}} + z\hat{\boldsymbol{z}}$ is the position vector of a geometric point in three-dimensional Euclidean space, $t$ is the time coordinate, and $v$ is the phase velocity for the propagation of waves inside the host medium. The linearity of the Helmholtz equation allows one to express its general solution as a superposition of these eigen-solutions — also known as plane-waves.

In the steady state, the time-dependence and the temporal frequency $\omega$ of the waves can be kept in the background by limiting the discussion to monochromatic (i.e., single-frequency) waves. If the EM waves happen to reside in free space, where their speed $v$ is the universal speed of light in vacuum, $c$, the aforementioned dispersion relation may be written as $k_x^2 + k_y^2 + k_z^2 = (2\pi/\lambda_0)^2$, with $\lambda_0 = 2\pi c/\omega$ being the vacuum wavelength of the (monochromatic) light under consideration. In what follows, we shall use the unit-vector $\boldsymbol{\sigma} = \sigma_x\hat{\boldsymbol{x}} + \sigma_y\hat{\boldsymbol{y}} + \sigma_z\hat{\boldsymbol{z}}$ to specify the direction of the $k$-vector, which is subsequently written as $\boldsymbol{k} = 2\pi\boldsymbol{\sigma}/\lambda_0$, thus ensuring that the dispersion relation is satisfied.

[‡] In situations where the vectorial nature of the EM fields can be ignored, the Helmholtz equation provides a good approximation to the Maxwell equations of classical electrodynamics. The plane-wave solutions of Maxwell's equations are described in Sec.6, where the connection between scalar and vector plane-waves becomes apparent.

Suppose the (scalar) light amplitude distribution $a_0(x,y)$, a complex function of the $x,y$ coordinates, is specified in the $xy$-plane at $z = 0$. The Fourier transform of this amplitude profile is

$$A_0(s_x, s_y) = \iint_{-\infty}^{\infty} a_0(x,y)\exp[-\mathrm{i}2\pi(s_x x + s_y y)]\mathrm{d}x\mathrm{d}y. \quad (5)$$

The inverse transform of $A_0(s_x, s_y)$ recovers the original amplitude distribution, as follows:

$$a_0(x,y) = a(x,y,z=0) = \iint_{-\infty}^{\infty} A_0(s_x, s_y)\exp[\mathrm{i}2\pi(s_x x + s_y y)]\mathrm{d}s_x\mathrm{d}s_y. \quad (6)$$

Assuming that the host medium is free space, and that the light is monochromatic with wavelength $\lambda_0$ (wave-number $k_0 = 2\pi/\lambda_0$), we now define $(\sigma_x, \sigma_y) = (\lambda_0 s_x, \lambda_0 s_y)$, and limit the range of integration to the unit circle in the $\sigma_x\sigma_y$-plane, as the points outside this circle correspond to evanescent waves, which exponentially decay with distance away from the source at $z = 0$.[4-7] Allowing each plane-wave to propagate from $z = 0$ to a parallel $xy$-plane located at $z$, we will have

$$a(x,y,z) = \lambda_0^{-2}\iint_{\text{unit circle}} A_0(\sigma_x/\lambda_0, \sigma_y/\lambda_0)\exp[\mathrm{i}(2\pi/\lambda_0)(\sigma_x x + \sigma_y y + \sigma_z z)]\,\mathrm{d}\sigma_x\mathrm{d}\sigma_y. \quad (7)$$

Letting the (fixed) observation point be $\boldsymbol{r} = x\hat{\boldsymbol{x}} + y\hat{\boldsymbol{y}} + z\hat{\boldsymbol{z}}$, and the unit-vector $\boldsymbol{\sigma}$ be defined as $\boldsymbol{\sigma} = \sigma_x\hat{\boldsymbol{x}} + \sigma_y\hat{\boldsymbol{y}} + \sigma_z\hat{\boldsymbol{z}}$, one can streamline Eq.(7), as follows:

$$a(\boldsymbol{r}) = \lambda_0^{-2}\iint_{\text{unit circle}} A_0(\sigma_x/\lambda_0, \sigma_y/\lambda_0)\exp(\mathrm{i}k_0\boldsymbol{\sigma}\cdot\boldsymbol{r})\,\mathrm{d}\sigma_x\mathrm{d}\sigma_y. \quad (8)$$

With reference to Fig.1(a), let the surface of a unit sphere centered at the origin of coordinates represent the location of the unit-vector $\boldsymbol{\sigma}$. The integral in Eq.(8) spans the entire surface of the hemisphere where $\sigma_z \geq 0$. Since, for any given observation point $\boldsymbol{r}$ in the system of Fig.1(a), the position vector $\boldsymbol{r}$ has a fixed orientation (as well as a fixed length), when $\boldsymbol{\sigma}$ revolves around $\boldsymbol{r}$ with a fixed angle $\varphi$, the traversed ring on the hemisphere's surface (depicted in Fig.1(b)) corresponds to a constant Fourier kernel $\exp(\mathrm{i}k_0\boldsymbol{\sigma}\cdot\boldsymbol{r}) = \exp(\mathrm{i}k_0 r\cos\varphi)$ in Eq.(8). Noting that the differential area on the hemisphere's surface times the obliquity factor $\sigma_z$ equals the projected differential area $\mathrm{d}\sigma_x\mathrm{d}\sigma_y$ onto the $xy$-plane, we define the auxiliary function $f(\varphi)$ as the average value of $\sigma_z A_0(\sigma_x/\lambda_0, \sigma_y/\lambda_0)$ over an entire ring of radius $\sin\varphi$ and infinitesimal width $\mathrm{d}\varphi$ —with the implicit understanding that $A_0 = 0$ wherever $\sigma_z < 0$. We will have

$$a(\boldsymbol{r}) = \lambda_0^{-2}\int_{\varphi=0}^{\pi} 2\pi\sin\varphi\, f(\varphi)\exp(\mathrm{i}k_0 r\cos\varphi)\,\mathrm{d}\varphi. \quad (9a)$$

In general, $f(\varphi)$ will be a slowly-varying function of $\varphi$, with $f(0) = \sigma_z A_0(\sigma_x/\lambda_0, \sigma_y/\lambda_0)$ evaluated at $\boldsymbol{\sigma} = \boldsymbol{r}/r = (x/r)\hat{\boldsymbol{x}} + (y/r)\hat{\boldsymbol{y}} + (z/r)\hat{\boldsymbol{z}}$, and $f(\pi) = 0$. With reference to Fig.1(a), the obliquity factor at $\varphi = 0$ may be written as $\sigma_z = z/r = \cos\theta$. Application of the method of integration by parts to Eq.(9a) now yields

$$a(\boldsymbol{r}) = \mathrm{i}(2\pi/k_0 r\lambda_0^2)\left[f(\varphi)\exp(\mathrm{i}k_0 r\cos\varphi)|_{\varphi=0}^{\pi} - \int_0^{\pi} f'(\varphi)\exp(\mathrm{i}k_0 r\cos\varphi)\,\mathrm{d}\varphi\right]. \quad (9b)$$

The remaining integral in Eq.(9b), evaluated via stationary-phase approximation,[2,23] turns out to be proportional to $(k_0 r)^{-1/2}$, which is negligible in the far field. All in all, the (complex) light amplitude distribution in the far field is found to be

$$a(x,y,z) = -\mathrm{i}[\cos\theta/(\lambda_0 r)]A_0[x/(\lambda_0 r), y/(\lambda_0 r)]\exp(\mathrm{i}k_0 r). \quad (10)$$

Aside from the curvature phase-factor $\exp(\mathrm{i}k_0 r)$ and the various scaling and normalization coefficients, the far field distribution in Eq.(10) is seen to be the Fourier transform of the initial distribution in the $xy$-plane at $z = 0$. On a spherical surface of radius $r$ centered at the origin of coordinates, the curvature phase-factor $\exp(\mathrm{i}k_0 r)$ is a constant. In this scalar version of the theory of diffraction, we have ignored the effects of polarization.[4] So long as the observation point is not too far away from the optical axis $z$ (i.e., the obliquity factor $\cos\theta$ is close to 1), the scalar theory proves to be highly accurate.[3,4] At large values of $\theta$, however, it is necessary to include the effects of polarization by introducing relevant obliquity factors for the various components of polarization.[22]

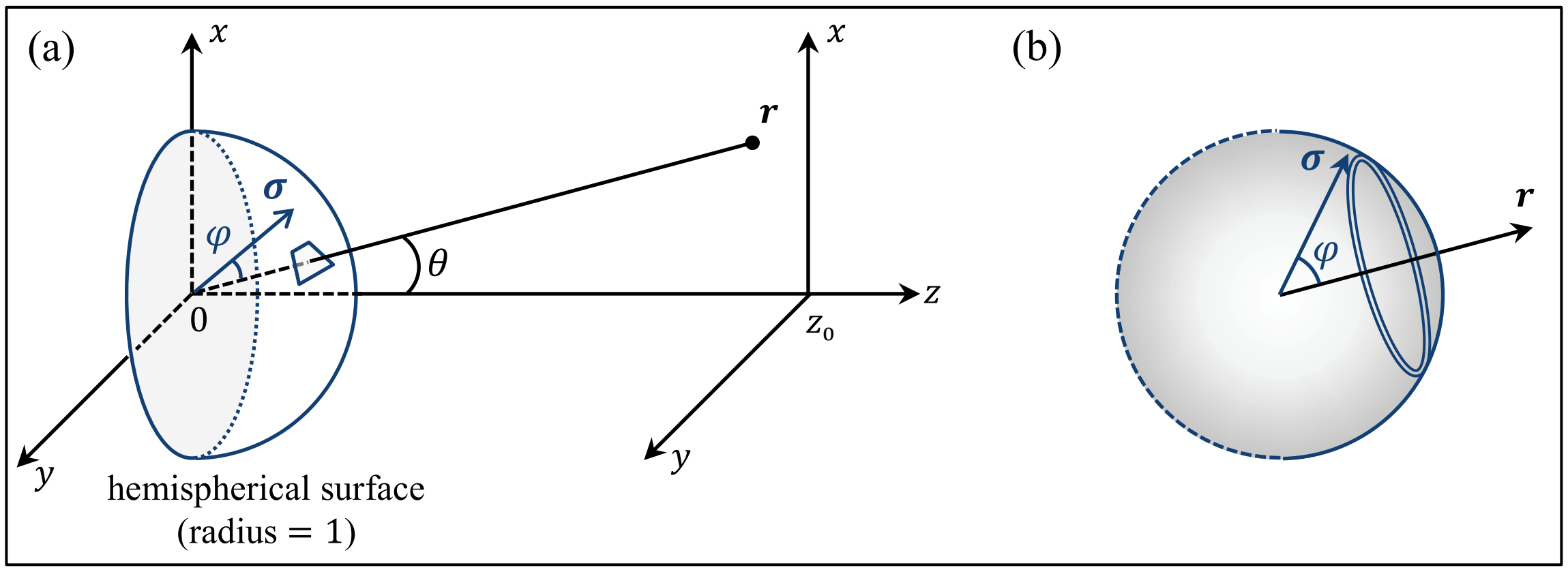

**Fig.1**. (a) An initial light amplitude distribution $a(x,y,z=0)$ in the $xy$-plane at $z = 0$ propagates in free space to arrive at a parallel $xy$-plane located at $z = z_0$. A hemispherical surface of radius 1.0 centered at the origin is the locus of the tips of the unit-vectors $\boldsymbol{\sigma} = \sigma_x\hat{\boldsymbol{x}} + \sigma_y\hat{\boldsymbol{y}} + \sigma_z\hat{\boldsymbol{z}}$ that define the propagation directions of the various plane-waves constituting the initial distribution. The observation point $\boldsymbol{r} = x\hat{\boldsymbol{x}} + y\hat{\boldsymbol{y}} + z\hat{\boldsymbol{z}}$ is an arbitrary point in space at which it is desired to compute the amplitude of the light beam. In the far field, where $k_0 r \gg 1$, the only plane-waves that significantly contribute to the observed light amplitude at $\boldsymbol{r}$ are those whose $\boldsymbol{\sigma}$-vectors fall within a small patch surrounding the point where the $\boldsymbol{r}$-vector pierces the hemisphere's surface. Within this patch, the angle $\varphi$ between $\boldsymbol{r}$ and $\boldsymbol{\sigma}$ is close to 0°. (b) For any given angle $\varphi$, the unit-vector $\boldsymbol{\sigma}$ describes a ring of radius $\sin\varphi$ circling $\boldsymbol{r}$ on the hemisphere's surface.

If the above argument is repeated for points on the left-hand side of the origin, where $z < 0$, one finds nearly the same relation as in Eq.(10) between the Fourier transform $A_0(s_x, s_y)$ of the field profile in the $xy$-plane at $z = 0$ and the incoming field amplitudes in locations to the left of the initial plane, the difference being that the curvature phase-factor now becomes $\exp(-\mathrm{i}k_0 r)$, the leading $-\mathrm{i}$ coefficient becomes $+\mathrm{i}$, and the $(x,y,z)$ coordinates of the observation point become $(-x,-y,-z)$. Figure 2 shows that a spherical wavefront on the left-hand side of the origin could represent the emergent beam from a lens whose focal plane is the $xy$-plane at $z = 0$. It is thus possible to consider $A_0[x/(\lambda_0 r), y/(\lambda_0 r)]$ as the light amplitude distribution over the convex spherical cap that touches the lens's second principal plane; that is, the scaled Fourier transform of the focal-plane distribution is directly related to the complex amplitude profile that emerges from the lens on this spherical cap. Needless to say, all the aberrations and other imperfections of the lens must be incorporated into the phase and amplitude profile on this (ideal) spherical wavefront that touches the second principal plane.[4]

If the lens is designed to bring the incident ray at height $\hbar$ on the first principal plane to the same height $\hbar$ on the emergent spherical wavefront, it is said to be an aplanatic lens. Such lenses satisfy Abbe's sine condition,[4,22] and their focal plane distribution, in accordance with Eq.(10), is the Fourier transform of the incident light distribution at their first principal plane.

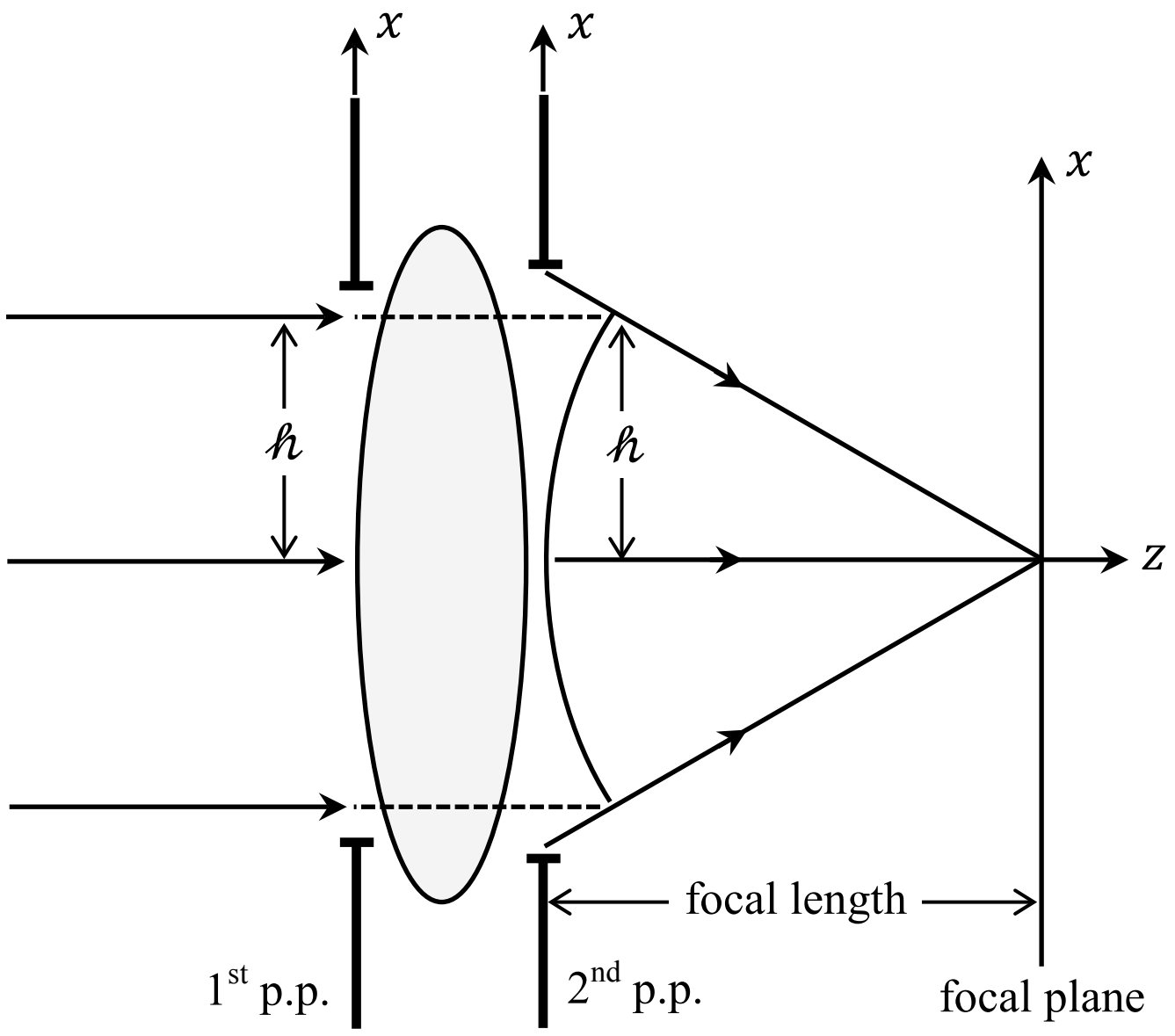

**Fig.2**. An aplanatic lens is designed to bring the rays that enter the pupil at the first principal plane onto the spherical surface of a convergent wavefront that emerges from the exit pupil at the second principal plane. All emerging rays, irrespective of their initial height $\hbar$ (i.e., distance from the $z$-axis), come together at the focal point, where the optical axis $z$ crosses the focal plane. The distance between the 2nd principal plane and the focal plane is the focal length of the lens.

**4. Eigenfunctions of the Fourier transform operator**. A number of ordinary functions are known to possess the special property that they are their own Fourier transforms.[1,23] Examples include $\exp(-\pi x^2)$, $1/\cosh(\pi x)$, $|x|^{-½}$, and $\text{comb}(x) = \sum_{n=-\infty}^{\infty} \delta(x-n)$, whose Fourier transforms are, respectively, $\exp(-\pi s^2)$, $1/\cosh(\pi s)$, $|s|^{-½}$, and $\text{comb}(s) = \sum_{n=-\infty}^{\infty} \delta(s-n)$. It is possible, however, to construct a large class of functions that, when Fourier transformed, reproduce the same function or a constant multiple thereof.[24] As an example, consider the even function $f(x)$, whose Fourier transform $F(s)$ is also even. Given that $\int_{-\infty}^{\infty} F(s)e^{-\text{i}2\pi xs}\text{d}s$ equals $f(-x)$, which is the same as $f(x)$, one can construct a new function $\varphi_+(x) = f(x) + F(x)$ whose Fourier transform is $\phi_+(s) = F(s) + f(s) = \varphi_+(s)$. In other words, $\varphi_+(x)$ is an eigenfunction of the Fourier transform operator with the eigenvalue of 1. Similarly, the function $\varphi_-(x) = f(x) - F(x)$, whose Fourier transform is $\phi_-(s) = F(s) - f(s) = -\varphi_-(s)$, is an eigenfunction with eigenvalue of $-1$.

As a second example, consider an odd function $g(x)$ whose Fourier transform $G(s)$ is also odd. Since $\int_{-\infty}^{\infty} G(s)e^{-\text{i}2\pi xs}\text{d}s = g(-x) = -g(x)$, both functions $\psi_\pm(x) = g(x) \pm \text{i}G(x)$ are eigenfunctions of the Fourier operator with respective eigenvalues of $\mp$i.

To prove that the only possible eigenvalues of the Fourier operator are $\pm 1$ and $\pm$i, suppose that $\Lambda(x)$ is an eigenfunction of the Fourier operator with eigenvalue $\lambda$. We will have

$$\int_{-\infty}^{\infty} \Lambda(x) \exp(-\text{i}2\pi sx)\, \text{d}x = \lambda\Lambda(s). \tag{11}$$

The inverse Fourier transform operation must yield

$$\Lambda(x) = \int_{-\infty}^{\infty} \lambda\Lambda(s) \exp(\text{i}2\pi sx)\, \text{d}s = \lambda^2\Lambda(-x). \tag{12}$$

Thus, it is necessary for $\Lambda(x)$ to be either even or odd, in which case $\lambda^2$ must be $+1$ or $-1$, respectively. When $\Lambda(x)$ is even, $\lambda$ will be $+1$ or $-1$, and when $\Lambda(x)$ is odd, $\lambda$ will be $+\mathrm{i}$ or $-\mathrm{i}$.

**5. Fourier transformation in electrodynamics**. In classical electrodynamics, one often treats Maxwell's equations as partial differential equations, attempting to find analytic solutions of these equations.[5] It is possible, however, to entirely avoid solving partial differential equations by taking Maxwell's equations into the Fourier domain, where everything is a superposition of plane-waves, for which differential equations turn into far simpler vector algebraic equations.[6] For example, the electric charge density $\rho(\boldsymbol{r},t)$, which is a scalar function of the spacetime $(\boldsymbol{r},t)$, can be written as a superposition of scalar plane-waves $\rho(\boldsymbol{k},\omega)\exp[\mathrm{i}(\boldsymbol{k}\cdot\boldsymbol{r}-\omega t)]$ in the four-dimensional Fourier space $(\boldsymbol{k},\omega)=(k_x,k_y,k_z,\omega)$. Similarly, the electric field $\boldsymbol{E}(\boldsymbol{r},t)$, a vector function of spacetime, may be written as a superposition of vector plane-waves in the form of $\boldsymbol{E}(\boldsymbol{k},\omega)\exp[\mathrm{i}(\boldsymbol{k}\cdot\boldsymbol{r}-\omega t)]$. Note that we are now using the same symbols $\rho$ and $\boldsymbol{E}$ to represent the functions in the $(\boldsymbol{r},t)$ spacetime domain as well as their transforms in the $(\boldsymbol{k},\omega)$ Fourier domain. (When the possibility of confusion exists, we shall specify the arguments of the function to indicate the domain over which the function is defined.) We have also absorbed the $2\pi$ factors appearing in Eqs.(1) and (2) as well as a minus sign into the various spatial and temporal frequency components, so that $(\boldsymbol{k},\omega)$ now stands for $2\pi(s_x,s_y,s_z,-s_t)$. With these modifications, a typical 4-dimensional Fourier transform and its inverse are written as follows:

$$\boldsymbol{E}(\boldsymbol{k},\omega)=\int_{-\infty}^{\infty}\boldsymbol{E}(\boldsymbol{r},t)\exp[-\mathrm{i}(\boldsymbol{k}\cdot\boldsymbol{r}-\omega t)]\,\mathrm{d}x\mathrm{d}y\mathrm{d}z\mathrm{d}t. \tag{13}$$

$$\boldsymbol{E}(\boldsymbol{r},t)=(2\pi)^{-4}\int_{-\infty}^{\infty}\boldsymbol{E}(\boldsymbol{k},\omega)\exp[\mathrm{i}(\boldsymbol{k}\cdot\boldsymbol{r}-\omega t)]\,\mathrm{d}k_x\mathrm{d}k_y\mathrm{d}k_z\mathrm{d}\omega. \tag{14}$$

There exist four sources and four fields in classical electrodynamics.[5-11] The sources are the free charge density $\rho_{\text{free}}(\boldsymbol{r},t)$, free current density $\boldsymbol{J}_{\text{free}}(\boldsymbol{r},t)$, electric dipole density (or polarization) $\boldsymbol{P}(\boldsymbol{r},t)$, and magnetic dipole density (or magnetization) $\boldsymbol{M}(\boldsymbol{r},t)$. The four fields are the electric field $\boldsymbol{E}(\boldsymbol{r},t)$, the magnetic field $\boldsymbol{H}(\boldsymbol{r},t)$, displacement $\boldsymbol{D}(\boldsymbol{r},t)$, and the magnetic induction $\boldsymbol{B}(\boldsymbol{r},t)$. In the international system of units ($SI$), the $\boldsymbol{D}$ and $\boldsymbol{B}$ fields are related to $\boldsymbol{E}$, $\boldsymbol{H}$, $\boldsymbol{P}$, and $\boldsymbol{M}$ in the following way:

$$\boldsymbol{D}(\boldsymbol{r},t)=\varepsilon_0\boldsymbol{E}(\boldsymbol{r},t)+\boldsymbol{P}(\boldsymbol{r},t). \tag{15}$$

$$\boldsymbol{B}(\boldsymbol{r},t)=\mu_0\boldsymbol{H}(\boldsymbol{r},t)+\boldsymbol{M}(\boldsymbol{r},t). \tag{16}$$

In these equations, $\varepsilon_0$ and $\mu_0$ are the permittivity and permeability of free space, which are related to the speed of light in vacuum via $c=(\mu_0\varepsilon_0)^{-½}$, and to the impedance of free space via $Z_0=(\mu_0/\varepsilon_0)^{½}$. The free charge and free current densities satisfy the charge-current continuity equation, namely,

$$\boldsymbol{\nabla}\cdot\boldsymbol{J}_{\text{free}}(\boldsymbol{r},t)+\partial_t\rho_{\text{free}}(\boldsymbol{r},t)=0. \tag{17}$$

The fields are related to their various sources through Maxwell's equations — a complete and consistent set of equations that govern the evolution of the fields for any given spatio-temporal source distribution. To determine the behavior of the sources in the presence of the fields, it is necessary to know the constitutive relations of the material media. The constitutive relation are rooted in the quantum mechanical structure of the media as well as the Lorentz force law and the dynamical equations of motion that govern the evolution of material bodies in response to various forces, torques, strains, and stresses.[4-11]

**6. Maxwell's equations**. Listed below are the four equations of Maxwell in their spatio-temporal differential form on the left, and in their Fourier domain representation on the right-hand side:

$$\boldsymbol{\nabla}\cdot\boldsymbol{D}(\boldsymbol{r},t)=\rho_{\text{free}}(\boldsymbol{r},t) \quad\rightarrow\quad \mathrm{i}\boldsymbol{k}\cdot\boldsymbol{D}(\boldsymbol{k},\omega)=\rho_{\text{free}}(\boldsymbol{k},\omega). \tag{18}$$

$$\boldsymbol{\nabla}\times\boldsymbol{H}(\boldsymbol{r},t)=\boldsymbol{J}_{\text{free}}(\boldsymbol{r},t)+\partial_t\boldsymbol{D}(\boldsymbol{r},t) \quad\rightarrow\quad \mathrm{i}\boldsymbol{k}\times\boldsymbol{H}(\boldsymbol{k},\omega)=\boldsymbol{J}_{\text{free}}(\boldsymbol{k},\omega)-\mathrm{i}\omega\boldsymbol{D}(\boldsymbol{k},\omega). \tag{19}$$

$$\boldsymbol{\nabla}\times\boldsymbol{E}(\boldsymbol{r},t)=-\partial_t\boldsymbol{B}(\boldsymbol{r},t) \quad\rightarrow\quad \mathrm{i}\boldsymbol{k}\times\boldsymbol{E}(\boldsymbol{k},\omega)=\mathrm{i}\omega\boldsymbol{B}(\boldsymbol{k},\omega). \tag{20}$$

$$\boldsymbol{\nabla}\cdot\boldsymbol{B}(\boldsymbol{r},t)=0 \quad\rightarrow\quad \mathrm{i}\boldsymbol{k}\cdot\boldsymbol{B}(\boldsymbol{k},\omega)=0. \tag{21}$$

If the sources $\rho_{\text{free}}$, $\boldsymbol{J}_{\text{free}}$, $\boldsymbol{P}$, and $\boldsymbol{M}$ are a priori specified, the Fourier domain algebraic equations (18)-(21) can be solved to yield the electromagnetic fields $\boldsymbol{E}$ and $\boldsymbol{H}$, as follows: Cross multiply Eq.(20) into $\boldsymbol{k}$ on the left-hand side, then substitute for $\boldsymbol{k}\times\boldsymbol{H}$ from Eq.(19) and for $\boldsymbol{k}\cdot\boldsymbol{E}$ from Eq.(18) to arrive at

$$\boldsymbol{k}\times(\boldsymbol{k}\times\boldsymbol{E})=\omega\boldsymbol{k}\times(\mu_0\boldsymbol{H}+\boldsymbol{M})$$

$$\rightarrow\quad (\boldsymbol{k}\cdot\boldsymbol{E})\boldsymbol{k}-(\boldsymbol{k}\cdot\boldsymbol{k})\boldsymbol{E}=\mu_0\omega(-\mathrm{i}\boldsymbol{J}_{\text{free}}-\varepsilon_0\omega\boldsymbol{E}-\omega\boldsymbol{P})+\omega\boldsymbol{k}\times\boldsymbol{M}$$

$$\rightarrow\quad (\boldsymbol{k}\cdot\boldsymbol{k}-\mu_0\varepsilon_0\omega^2)\boldsymbol{E}=\varepsilon_0^{-1}(-\mathrm{i}\rho_{\text{free}}-\boldsymbol{k}\cdot\boldsymbol{P})\boldsymbol{k}+\mathrm{i}\mu_0\omega\boldsymbol{J}_{\text{free}}+\mu_0\omega^2\boldsymbol{P}-\omega\boldsymbol{k}\times\boldsymbol{M}$$

$$\rightarrow\quad \boldsymbol{E}(\boldsymbol{k},\omega)=\frac{\mathrm{i}\mu_0\omega(\boldsymbol{J}_{\text{free}}-\mathrm{i}\omega\boldsymbol{P}+\mathrm{i}\mu_0^{-1}\boldsymbol{k}\times\boldsymbol{M})-\mathrm{i}\varepsilon_0^{-1}(\rho_{\text{free}}-\mathrm{i}\boldsymbol{k}\cdot\boldsymbol{P})\boldsymbol{k}}{k^2-(\omega/c)^2}. \tag{22}$$

The vector identity $\boldsymbol{a}\times(\boldsymbol{b}\times\boldsymbol{c})=(\boldsymbol{a}\cdot\boldsymbol{c})\boldsymbol{b}-(\boldsymbol{a}\cdot\boldsymbol{b})\boldsymbol{c}$ has been used in this derivation. The $H$-field may now be obtained by substituting $\boldsymbol{E}(\boldsymbol{k},\omega)$ from Eq.(22) into Eq.(20), which yields

$$\boldsymbol{H}(\boldsymbol{k},\omega)=\frac{\mathrm{i}\boldsymbol{k}\times(\boldsymbol{J}_{\text{free}}-\mathrm{i}\omega\boldsymbol{P}+\mathrm{i}\mu_0^{-1}\boldsymbol{k}\times\boldsymbol{M})}{k^2-(\omega/c)^2}-\mu_0^{-1}\boldsymbol{M}. \tag{23}$$

The EM fields in the spacetime domain are straightforwardly computed by inverse Fourier transformation of $\boldsymbol{E}(\boldsymbol{k},\omega)$ and $\boldsymbol{H}(\boldsymbol{k},\omega)$. The singularity at $k=\omega/c$ needs careful attention, as it corresponds to plane-waves that could reside in free space in the absence of all sources. One typically handles such singularities by giving $\omega$ a small imaginary value that would then be driven to zero at the end of the calculation.[5,6] Alternatively, one may keep $\omega$ real at the expense of introducing an additional superposition of free-space plane-waves in the end that would restore the expected physical behavior of the overall solution. A specific example of the latter approach to solving Maxwell's equations appears in Appendix A, in relation to the all-encompassing Example presented at the end of the following section.

Note that Maxwell's fourth equation, Eq.(21), has not been used in any of the above derivations, the reason being that Eq.(21) is implicit in Eq.(20), which can be seen by dot-multiplying Eq.(20) into $\boldsymbol{k}$ to arrive at

$$\boldsymbol{k}\cdot(\boldsymbol{k}\times\boldsymbol{E})=\omega\boldsymbol{k}\cdot\boldsymbol{B} \quad\rightarrow\quad \omega\boldsymbol{k}\cdot\boldsymbol{B}=(\boldsymbol{k}\times\boldsymbol{k})\cdot\boldsymbol{E}=0. \tag{24}$$

Thus, the only time that Maxwell's fourth equation needs to be explicitly taken into account is when $\omega=0$, in which case Eq.(24) does not guarantee the vanishing of $\boldsymbol{k}\cdot\boldsymbol{B}$.

**7. Standard scalar and vector potentials.**[4-11] The vector potential $\boldsymbol{A}(\boldsymbol{r},t)$ is defined as a vector field whose curl equals the magnetic induction $\boldsymbol{B}(\boldsymbol{r},t)$. The relation is written below, first in its spacetime version, and then in the Fourier domain:

$$\boldsymbol{B}(\boldsymbol{r},t)=\boldsymbol{\nabla}\times\boldsymbol{A}(\boldsymbol{r},t) \quad\rightarrow\quad \boldsymbol{B}(\boldsymbol{k},\omega)=\mathrm{i}\boldsymbol{k}\times\boldsymbol{A}(\boldsymbol{k},\omega). \tag{25}$$

This ensures the satisfaction of Maxwell's 4th equation due to the fact that $\boldsymbol{k}\cdot\boldsymbol{B}(\boldsymbol{k},\omega)$, being equal to $\mathrm{i}(\boldsymbol{k}\times\boldsymbol{k})\cdot\boldsymbol{A}(\boldsymbol{k},\omega)$, is guaranteed to vanish. With the introduction of the vector potential, Maxwell's 3rd equation can be written as $\boldsymbol{\nabla}\times(\boldsymbol{E}+\partial_t\boldsymbol{A})=0$, which, in the Fourier domain, becomes $\mathrm{i}\boldsymbol{k}\times(\boldsymbol{E}-\mathrm{i}\omega\boldsymbol{A})=0$. This indicates that the vector $\mathrm{i}(\boldsymbol{E}-\mathrm{i}\omega\boldsymbol{A})$ must be aligned with $\boldsymbol{k}$, or, put differently, a scalar coefficient $\psi$ must exist such that $\mathrm{i}(\boldsymbol{E}-\mathrm{i}\omega\boldsymbol{A})=\psi\boldsymbol{k}$. Needless to say, the scalar coefficient $\psi$ must be a function of $(\boldsymbol{k},\omega)$, and the resulting formula gives the electric field in terms of the scalar and vector potentials $\psi$ and $\boldsymbol{A}$, as

$$\boldsymbol{E}(\boldsymbol{k},\omega)=-\mathrm{i}\boldsymbol{k}\psi(\boldsymbol{k},\omega)+\mathrm{i}\omega\boldsymbol{A}(\boldsymbol{k},\omega)\quad\rightarrow\quad\boldsymbol{E}(\boldsymbol{r},t)=-\boldsymbol{\nabla}\psi(\boldsymbol{r},t)-\partial_t\boldsymbol{A}(\boldsymbol{r},t).\qquad(26)$$

With the above definitions, the scalar and vector potentials as well as their relations with the $\boldsymbol{E}$ and $\boldsymbol{B}$ fields are established. Note, however, that the component of $\boldsymbol{A}(\boldsymbol{k},\omega)$ along the direction of the $\boldsymbol{k}$-vector is not specified at all. All that Eq.(25) does is fix the component of $\boldsymbol{A}(\boldsymbol{k},\omega)$ in a direction perpendicular to $\boldsymbol{k}$. One remains free to choose the projection of $\boldsymbol{A}(\boldsymbol{k},\omega)$ along the $\boldsymbol{k}$ vector. This is called the freedom to set the gauge,[5-11] and in what follows, we shall exercise this freedom in a particular way to simplify the relations between the potentials and the sources. (This choice of gauge is equivalent to being free to choose the divergence of $\boldsymbol{A}$ in the spacetime domain, which is the common, textbook way the magnetic vector potential is derived.)

Substituting the $\boldsymbol{E}$ and $\boldsymbol{B}$ fields of Eqs.(25) and (26) into Maxwell's 2$^{\text{nd}}$ equation, we find

$$\mathrm{i}\boldsymbol{k}\times\mu_0^{-1}(\mathrm{i}\boldsymbol{k}\times\boldsymbol{A}-\boldsymbol{M})=\boldsymbol{J}_{\text{free}}-\mathrm{i}\omega[\varepsilon_0(-\mathrm{i}\boldsymbol{k}\psi+\mathrm{i}\omega\boldsymbol{A})+\boldsymbol{P}]$$

$$\rightarrow\quad[\boldsymbol{k}\cdot\boldsymbol{k}-(\omega/c)^2]\boldsymbol{A}=\mu_0(\boldsymbol{J}_{\text{free}}-\mathrm{i}\omega\boldsymbol{P}+\mu_0^{-1}\mathrm{i}\boldsymbol{k}\times\boldsymbol{M})+[\boldsymbol{k}\cdot\boldsymbol{A}-(\omega/c^2)\psi]\boldsymbol{k}.\qquad(27)$$

Invoking the freedom to set the gauge, we now remove the last term on the right-hand side of Eq.(27). This is the so-called Lorenz gauge, which, expressed in the Fourier domain as well as the spacetime domain, can be written as

$$\boldsymbol{k}\cdot\boldsymbol{A}(\boldsymbol{k},\omega)-(\omega/c^2)\psi(\boldsymbol{k},\omega)=0\qquad\rightarrow\qquad\boldsymbol{\nabla}\cdot\boldsymbol{A}(\boldsymbol{r},t)+(1/c^2)\partial_t\psi(\boldsymbol{r},t)=0.\qquad(28)$$

In the Lorenz gauge, a simplified Eq.(27) yields

$$\boldsymbol{A}(\boldsymbol{k},\omega)=\frac{\mu_0(\boldsymbol{J}_{\text{free}}-\mathrm{i}\omega\boldsymbol{P})+\mathrm{i}\boldsymbol{k}\times\boldsymbol{M}}{k^2-(\omega/c)^2}.\qquad(29)$$

Finally, substitution of Eqs.(26) and (28) into Maxwell's first equation yields the scalar potential (again, in the Lorenz gauge), as follows:

$$\psi(\boldsymbol{k},\omega)=\frac{\rho_{\text{free}}-\mathrm{i}\boldsymbol{k}\cdot\boldsymbol{P}}{\varepsilon_0[k^2-(\omega/c)^2]}.\qquad(30)$$

Once the potentials are obtained in the Fourier domain, they can be returned to the spacetime domain via inverse Fourier transformation. The $\boldsymbol{E}$ and $\boldsymbol{B}$ fields are then computed from Eqs.(25) and (26). Alternatively, one may choose to first compute the $\boldsymbol{E}$ and $\boldsymbol{B}$ fields in the Fourier domain (in which case one would obtain Eqs.(22) and (23) of the preceding section), then carry out the inverse Fourier transforms.

Solving Maxwell's equations in the $(\boldsymbol{k},\omega)$ domain is algebraically straightforward; it also provides a degree of flexibility in arranging the sources, the fields, and the potentials in various groups and in different combinations that is not easy to achieve otherwise (i.e., when dealing with differential equations in the spacetime domain). While there exist specific problems that may be easier to solve in the $(\boldsymbol{r},t)$ space, and others for which operating in the $(\boldsymbol{k},\omega)$ domain would be advantageous, there is no denying the appeal of the Fourier domain methods for their conceptual clarity and operational flexibility as well as simplicity. In light of the arguments presented in this and the preceding section, it should not be difficult to see, for instance, that in the absence of $\rho_{\text{free}}$ and $\boldsymbol{J}_{\text{free}}$, one could define the scalar and vector potentials in an entirely different way, such that $\boldsymbol{D}=\boldsymbol{\nabla}\times\boldsymbol{A}$ and $\boldsymbol{H}=\boldsymbol{\nabla}\psi+\partial_t\boldsymbol{A}$. Having thus satisfied Maxwell's 1$^{\text{st}}$ and 2nd equations, one proceeds to set a gauge and then to relate the new $\boldsymbol{A}$ and $\psi$ via Maxwell's 3rd and 4th equations to the remaining sources $\boldsymbol{P}$ and $\boldsymbol{M}$. The resulting formulas will look different than Eqs.(29) and (30), and, depending on the situation under investigation, they may provide an easier path to the solution for the EM fields. Nevertheless, when all is said and done, the new potentials will yield precisely the same solutions for the desired EM fields as do the old ones.

**Example**. As an application of the method described in the preceding sections, we examine the case of EM radiation by the infinite sheet of oscillating current depicted in Fig.3. The electric current density, confined to a thin sheet in the $xy$-plane, is specified as

$$\boldsymbol{J}_{\text{free}}(\boldsymbol{r},t) = J_{s0}\cos(k_{y0}y - \omega_0 t)\,\delta(z)\hat{\boldsymbol{x}}. \tag{31}$$

Here, the surface current amplitude $J_{s0}$, the spatial frequency $k_{y0}$ along the $y$-axis, and the oscillation frequency $\omega_0$ are constants. Since $\boldsymbol{\nabla}\cdot\boldsymbol{J}_{\text{free}} = 0$, the charge-current continuity equation confirms that no associated free charges reside in the system, that is, $\rho_{\text{free}}(\boldsymbol{r},t) = 0$. Given that $\boldsymbol{P}(\boldsymbol{r},t)$ and $\boldsymbol{M}(\boldsymbol{r},t)$ are also absent in this example, there will be no scalar potential and, therefore, one only needs to compute the vector potential $\boldsymbol{A}(\boldsymbol{r},t)$, from which the radiated $\boldsymbol{E}$ and $\boldsymbol{H}$ fields will subsequently be determined. We begin by transforming the current density to the Fourier domain; that is,

$$\boldsymbol{J}_{\text{free}}(\boldsymbol{k},\omega) = \int_{-\infty}^{\infty} J_{s0}\hat{\boldsymbol{x}}\cos(k_{y0}y - \omega_0 t)\,\delta(z)\exp[-\mathrm{i}(\boldsymbol{k}\cdot\boldsymbol{r} - \omega t)]\,\mathrm{d}x\mathrm{d}y\mathrm{d}z\mathrm{d}t$$

$$= 4\pi^3 J_{s0}\hat{\boldsymbol{x}}\,[\delta(k_x)\delta(k_y - k_{y0})\delta(\omega - \omega_0) + \delta(k_x)\delta(k_y + k_{y0})\delta(\omega + \omega_0)]. \tag{32}$$

The vector potential is thus given by

$$\boldsymbol{A}(\boldsymbol{k},\omega) = \frac{\mu_0 \boldsymbol{J}_{\text{free}}}{k^2 - (\omega/c)^2} = \frac{4\pi^3\mu_0 J_{s0}\hat{\boldsymbol{x}}\,[\delta(k_x)\delta(k_y - k_{y0})\delta(\omega-\omega_0) + \delta(k_x)\delta(k_y + k_{y0})\delta(\omega+\omega_0)]}{k_x^2 + k_y^2 + k_z^2 - (\omega/c)^2}. \tag{33}$$

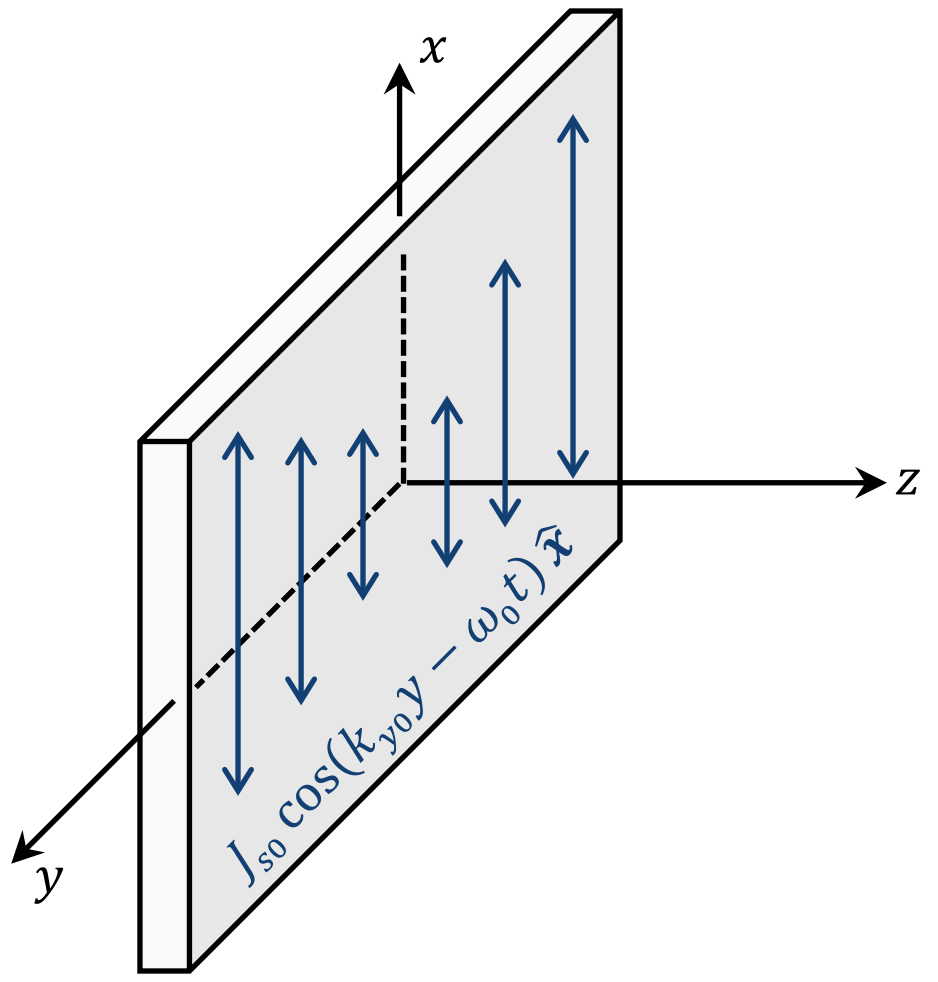

**Fig.3**. An infinite sheet in the $xy$-plane carries the current-density $\boldsymbol{J}_{\text{free}}(\boldsymbol{r},t) = J_{s0}\cos(k_{y0}y - \omega_0 t)\,\delta(z)\hat{\boldsymbol{x}}$. When $|k_{y0}| < \omega_0/c$, two plane-waves, one propagating on the right-, the other on the left-hand side of the sheet, emanate into the surrounding free space. In the case of $|k_{y0}| > \omega_0/c$, exponentially-decaying evanescent fields hug the current-carrying sheet from both sides.

One can readily verify that $\boldsymbol{k}\cdot\boldsymbol{A}=0$, and that, therefore, the Lorenz gauge is satisfied. The inverse transform of $\boldsymbol{A}(\boldsymbol{k},\omega)$ is found to be (see Appendix A for derivation details):

$$\boldsymbol{A}(\boldsymbol{r},t)=(2\pi)^{-4}\int_{-\infty}^{\infty}\boldsymbol{A}(\boldsymbol{k},\omega)\exp[\mathrm{i}(\boldsymbol{k}\cdot\boldsymbol{r}-\omega t)]\,\mathrm{d}k_x\mathrm{d}k_y\mathrm{d}k_z\mathrm{d}\omega$$

$$=\frac{\mu_0 J_{s0}\hat{\boldsymbol{x}}}{2k_{z0}}\begin{cases}-\sin(k_{y0}y+k_{z0}|z|-\omega_0 t); & |k_{y0}|<\omega_0/c\\ e^{-k_{z0}|z|}\cos(k_{y0}y-\omega_0 t); & |k_{y0}|>\omega_0/c.\end{cases}\tag{34}$$

Here $k_{z0}=\sqrt{(\omega_0/c)^2-k_{y0}^2}$ when $|k_{y0}|<\omega_0/c$, and $k_{z0}=\sqrt{k_{y0}^2-(\omega_0/c)^2}$ when $|k_{y0}|>\omega_0/c$. The radiated electric and magnetic fields are derived from the vector potential, as follows:

$$\boldsymbol{E}(\boldsymbol{r},t)=-\partial_t\boldsymbol{A}(\boldsymbol{r},t)=-\frac{\mu_0\omega_0 J_{s0}\hat{\boldsymbol{x}}}{2k_{z0}}\begin{cases}\cos(k_{y0}y+k_{z0}|z|-\omega_0 t); & |k_{y0}|<\omega_0/c,\\ e^{-k_{z0}|z|}\sin(k_{y0}y-\omega_0 t); & |k_{y0}|>\omega_0/c.\end{cases}\tag{35}$$

$$\boldsymbol{H}(\boldsymbol{r},t)=\mu_0^{-1}\boldsymbol{\nabla}\times\boldsymbol{A}(\boldsymbol{r},t)$$

$$=-\frac{J_{s0}}{2k_{z0}}\begin{cases}[\mathrm{sign}(z)k_{z0}\hat{\boldsymbol{y}}-k_{y0}\hat{\boldsymbol{z}}]\cos(k_{y0}y+k_{z0}|z|-\omega_0 t); & |k_{y0}|<\omega_0/c,\\ e^{-k_{z0}|z|}[\mathrm{sign}(z)k_{z0}\hat{\boldsymbol{y}}\cos(k_{y0}y-\omega_0 t)-k_{y0}\hat{\boldsymbol{z}}\sin(k_{y0}y-\omega_0 t)]; & |k_{y0}|>\omega_0/c.\end{cases}\tag{36}$$

These are the complete solutions of Maxwell's equations for radiation from the current-carrying sheet depicted in Fig.3. When $|k_{y0}|<\omega_0/c$, two plane-waves emerge from the sheet, one propagating along $k_{y0}\hat{\boldsymbol{y}}+k_{z0}\hat{\boldsymbol{z}}$ on the right, the other along $k_{y0}\hat{\boldsymbol{y}}-k_{z0}\hat{\boldsymbol{z}}$ on the left-hand side. In the case of $|k_{y0}|>\omega_0/c$, exponentially decaying evanescent waves reside on both sides of the sheet.[4-7]

At the surface of the sheet ($z=0^{\pm}$) Maxwell's boundary conditions are seen to be satisfied. In particular, the discontinuity of the tangential component $H_y$ of the magnetic field precisely matches the surface current density $J_{s0}\cos(k_{y0}y-\omega_0 t)$. The $E$-field at $z=0$ is responsible for radiation resistance.[5-11] It is not difficult to verify that the rate of radiation of EM energy (per unit area per unit time) is equal to the rate at which this radiation reaction $E$-field extracts energy from the surface current. Similarly, it is easily confirmed that the rate of expulsion of EM momentum is consistent with the self force exerted by the action of the magnetic field $B_z$ on the surface current. To see these, observe that the time-averaged Poynting vector is given by[4-11]

$$\langle\boldsymbol{S}(\boldsymbol{r},t)\rangle=\langle\boldsymbol{E}(\boldsymbol{r},t)\times\boldsymbol{H}(\boldsymbol{r},t)\rangle=\frac{\mu_0\omega_0 J_{s0}^2}{8k_{z0}^2}\begin{cases}k_{y0}\hat{\boldsymbol{y}}+\mathrm{sign}(z)k_{z0}\hat{\boldsymbol{z}}; & |k_{y0}|<\omega_0/c,\\ \exp(-2k_{z0}|z|)\,k_{y0}\hat{\boldsymbol{y}}; & |k_{y0}|>\omega_0/c.\end{cases}\tag{37}$$

Note that the radiated energy per unit area per unit time is equal to $\langle\boldsymbol{E}(\boldsymbol{r},t)\cdot\boldsymbol{J}(\boldsymbol{r},t)\rangle$ (integrated over the sheet thickness), which is readily found from Eqs.(31) and (35) to be $-\mu_0\omega_0 J_{s0}^2/(4k_{z0})$. The minus sign indicates that the energy is being *extracted* from the current source by the action of the self-field. Needless to say, the radiated energy density in Eq.(37) must be multiplied by 2 to account for the fact that plane-waves emerge on both sides of the current-carrying sheet, and by the obliquity factor $k_{z0}/k_0$ to account for the footprints of the emergent plane-waves on the $xy$-plane of the sheet.

The rate of radiation of EM momentum is given by the EM momentum density, $\boldsymbol{S}(\boldsymbol{r},t)/c^2$, times the speed of light $c$ in vacuum.[5-7] Accounting for both plane-waves, Eq.(37) shows that the net rate of radiation of EM momentum is $\mu_0\omega_0 J_{s0}^2 k_{y0}\hat{\boldsymbol{y}}/(4k_{z0}^2 c)$. The Lorentz force exerted by the self-field per unit area of the current-carrying sheet is $\langle\boldsymbol{J}(\boldsymbol{r},t)\times\mu_0\boldsymbol{H}(\boldsymbol{r},t)\rangle$ (integrated over the sheet thickness), which is seen from Eqs.(31) and (36) to be $-\mu_0 J_{s0}^2 k_{y0}\hat{\boldsymbol{y}}/(4k_{z0})$. The minus sign indicates that the exerted force is opposite in direction to the emitted EM momentum, which is necessary if the total momentum of the system is to be conserved. Accounting once again for the obliquity factor $k_{z0}/k_0$, the above argument shows that the overall momentum of the system is in fact conserved.

**8. Hertz vector potentials**. When solving electrodynamics problems, it is occasionally convenient to introduce the Hertz vector potentials, with the aid of which the $\boldsymbol{D}$, $\boldsymbol{B}$, $\boldsymbol{E}$, and $\boldsymbol{H}$ fields are subsequently determined.[4] Returning to Maxwell's equations in the Fourier domain, in the absence of free charges and currents, that is, when $\rho_{\text{free}}(\boldsymbol{k},\omega)=0$ and $\boldsymbol{J}_{\text{free}}(\boldsymbol{k},\omega)=0$, one manipulates the 2nd equation (using substitutions from the 3rd and 4th equations) to arrive at

$$\boldsymbol{k}\times[\boldsymbol{k}\times\boldsymbol{B}(\boldsymbol{k},\omega)]=-\mu_0\omega\boldsymbol{k}\times\boldsymbol{D}(\boldsymbol{k},\omega)+\boldsymbol{k}\times[\boldsymbol{k}\times\boldsymbol{M}(\boldsymbol{k},\omega)]$$

$$\rightarrow\quad \boldsymbol{B}(\boldsymbol{k},\omega)=\mu_0\omega\boldsymbol{k}\times\frac{\boldsymbol{P}(\boldsymbol{k},\omega)}{k^2-(\omega/c)^2}-\boldsymbol{k}\times\left[\boldsymbol{k}\times\frac{\boldsymbol{M}(\boldsymbol{k},\omega)}{k^2-(\omega/c)^2}\right]; \tag{38}$$

see Appendix B for derivation details. The Hertz potentials may now be defined in the Fourier domain, as follows:

$$\boldsymbol{\Pi}_{\text{e}}(\boldsymbol{k},\omega)=\frac{\boldsymbol{P}(\boldsymbol{k},\omega)}{k^2-(\omega/c)^2}\qquad\text{and}\qquad\boldsymbol{\Pi}_{\text{m}}(\boldsymbol{k},\omega)=\frac{\boldsymbol{M}(\boldsymbol{k},\omega)}{k^2-(\omega/c)^2}. \tag{39}$$

Consequently,

$$\boldsymbol{B}(\boldsymbol{r},t)=\boldsymbol{\nabla}\times[\boldsymbol{\nabla}\times\boldsymbol{\Pi}_{\text{m}}(\boldsymbol{r},t)+\mu_0\partial_t\boldsymbol{\Pi}_{\text{e}}(\boldsymbol{r},t)]. \tag{40}$$

In similar fashion, Maxwell's 3rd equation can be manipulated (using substitutions from the 1st and 2nd equations) to obtain

$$\boldsymbol{k}\times[\boldsymbol{k}\times\boldsymbol{D}(\boldsymbol{k},\omega)]=\varepsilon_0\omega\boldsymbol{k}\times\boldsymbol{B}(\boldsymbol{k},\omega)+\boldsymbol{k}\times[\boldsymbol{k}\times\boldsymbol{P}(\boldsymbol{k},\omega)]$$

$$\rightarrow\quad \boldsymbol{D}(\boldsymbol{k},\omega)=-\varepsilon_0\omega\boldsymbol{k}\times\frac{\boldsymbol{M}(\boldsymbol{k},\omega)}{k^2-(\omega/c)^2}-\boldsymbol{k}\times\left[\boldsymbol{k}\times\frac{\boldsymbol{P}(\boldsymbol{k},\omega)}{k^2-(\omega/c)^2}\right]; \tag{41}$$

see Appendix B for derivation details. Consequently,

$$\boldsymbol{D}(\boldsymbol{r},t)=\boldsymbol{\nabla}\times[\boldsymbol{\nabla}\times\boldsymbol{\Pi}_{\text{e}}(\boldsymbol{r},t)-\varepsilon_0\partial_t\boldsymbol{\Pi}_{\text{m}}(\boldsymbol{r},t)]. \tag{42}$$

Applying inverse Fourier transformation to Eqs.(39) reveals the Hertz potentials as solutions of the following (inhomogeneous) Helmholtz equations[4] in the spacetime domain:

$$\left(\nabla^2-\frac{1}{c^2}\frac{\partial^2}{\partial t^2}\right)\boldsymbol{\Pi}_{\text{e}}(\boldsymbol{r},t)=-\boldsymbol{P}(\boldsymbol{r},t), \tag{43a}$$

$$\left(\nabla^2-\frac{1}{c^2}\frac{\partial^2}{\partial t^2}\right)\boldsymbol{\Pi}_{\text{m}}(\boldsymbol{r},t)=-\boldsymbol{M}(\boldsymbol{r},t). \tag{43b}$$

The above equations can be shown to have the following solutions:[4-7]

$$\boldsymbol{\Pi}_{\text{e}}(\boldsymbol{r},t)=\iiint_{-\infty}^{\infty}\frac{\boldsymbol{P}(\boldsymbol{r}',\ t-|\boldsymbol{r}-\boldsymbol{r}'|/c)}{4\pi|\boldsymbol{r}-\boldsymbol{r}'|}\mathrm{d}\boldsymbol{r}', \tag{44a}$$

$$\boldsymbol{\Pi}_{\text{m}}(\boldsymbol{r},t)=\iiint_{-\infty}^{\infty}\frac{\boldsymbol{M}(\boldsymbol{r}',\ t-|\boldsymbol{r}-\boldsymbol{r}'|/c)}{4\pi|\boldsymbol{r}-\boldsymbol{r}'|}\mathrm{d}\boldsymbol{r}'. \tag{44b}$$

The connection between the Hertz vector potentials $(\boldsymbol{\Pi}_{\rm e}, \boldsymbol{\Pi}_{\rm m})$ and the standard scalar and vector potentials $(\psi, \boldsymbol{A})$ is readily seen through a comparison of Eq.(39) with Eqs.(29) and (30), keeping in mind that the Hertz potentials are defined in the absence of $\rho_{\rm free}$ and $\boldsymbol{J}_{\rm free}$. In solving electrodynamics problems, it is sometimes convenient to find the Hertz potentials at first, then derive the $\boldsymbol{B}$ and $\boldsymbol{D}$ fields from Eqs.(40) and (42). With $\boldsymbol{B}(\boldsymbol{r},t)$ and $\boldsymbol{D}(\boldsymbol{r},t)$ at hand, given that $\boldsymbol{P}(\boldsymbol{r},t)$ and $\boldsymbol{M}(\boldsymbol{r},t)$ are already specified, finding the $\boldsymbol{E}$ and $\boldsymbol{H}$ fields is rather trivial.

**9. Fourier transformation in quantum optics**. There exist numerous situations where the Fourier transforms appear naturally in quantum mechanics and quantum optics.[12,13] Here, we present a single example that shows a beautiful property of the wave-function of a particle trapped in the quadratic potential of a harmonic oscillator. In the mass-and-spring model of the one-dimensional harmonic oscillator, a particle of mass $m$, attached to a spring (spring constant $= \kappa$), oscillates back and forth along the $x$-axis. The Schrödinger equation describing the particle's wave-function is[12]

$$\mathrm{i}\hbar \frac{\partial}{\partial t}\psi(x,t) = -\left(\frac{\hbar^2}{2m}\right)\frac{\partial^2}{\partial x^2}\psi(x,t) + \tfrac{1}{2}\kappa x^2\psi(x,t). \tag{45}$$

The energy eigen-modes $\psi_n(x)e^{-\mathrm{i}(\mathcal{E}_n/\hbar)t}$ satisfy the time-independent Schrödinger equation,

$$\psi_n''(x) + (m/\hbar^2)(2\mathcal{E}_n - \kappa x^2)\psi_n(x) = 0. \tag{46}$$

Guessing that $\psi_n(x) = \xi_n(\alpha x)e^{-\beta x^2}$ is the general form of the solution—with the function $\xi_n(\cdot)$ and the constant parameters $\alpha$ and $\beta$ as yet unspecified—substitution into Eq.(46) yields

$$\alpha^2\xi_n''(\alpha x) - 4\alpha\beta x\xi_n'(\alpha x) + 2\left(\frac{m\mathcal{E}_n}{\hbar^2} - \beta\right)\xi_n(\alpha x) + \left(4\beta^2 - \frac{m\kappa}{\hbar^2}\right)x^2\xi_n(\alpha x) = 0. \tag{47}$$

Setting $\beta = \sqrt{m\kappa}/2\hbar$ eliminates the last term on the left-hand side of Eq.(47), leaving us with

$$\xi_n''(\alpha x) - 2\left(\frac{2\beta}{\alpha^2}\right)\alpha x\xi_n'(\alpha x) + 2\left(\frac{m\mathcal{E}_n}{\alpha^2\hbar^2} - \frac{\beta}{\alpha^2}\right)\xi_n(\alpha x) = 0. \tag{48}$$

This reduces to Hermite's equation if one sets $\alpha = \sqrt{2\beta}$ and $\mathcal{E}_n = 2\beta\hbar^2(n + \tfrac{1}{2})/m$. (See Appendix C for the properties of Hermite's equation and the corresponding polynomials.) Defining the resonance frequency of the mass-and-spring system as $\omega_0 = \sqrt{\kappa/m}$, we now have $\alpha = \sqrt{m\omega_0/\hbar}$ and $\mathcal{E}_n = (n + \tfrac{1}{2})\hbar\omega_0$. The general eigen-mode solution for the Schrödinger wave-function is thus given by

$$\psi_n(x,t) = \left(\frac{\alpha}{\sqrt{\pi}\, n!2^n}\right)^{\frac{1}{2}} H_n(\alpha x)e^{-\frac{1}{2}(\alpha x)^2 - \mathrm{i}(\mathcal{E}_n/\hbar)t}, \tag{49}$$

where the upfront normalization coefficient ensures that $\int_{-\infty}^{\infty}|\psi_n(x,t)|^2\mathrm{d}x = 1$. A general superposition of the various eigen-modes of the harmonic oscillator is written as

$$\psi(x,t) = e^{-\frac{1}{2}\mathrm{i}\omega_0 t}\sum_n c_n\psi_n(x)e^{-\mathrm{i}n\omega_0 t}. \tag{50}$$

Aside from the constant phase-factor $e^{-\frac{1}{2}\mathrm{i}\omega_0 t}$, this is a periodic function of time, with period $T = 2\pi/\omega_0$. When the time $t$ advances by a quarter of one period $T$, the phase of the $n^{\rm th}$ eigen-mode changes by $e^{-\frac{1}{2}\mathrm{i}n\pi} = (-\mathrm{i})^n$. Considering that the Fourier transform of $\psi(x,t)$ with respect to its space coordinate $x$ at any given time $t$ is given by

$$\int_{-\infty}^{\infty}\psi(x,t)\exp(-\mathrm{i}2\pi sx)\,\mathrm{d}x = e^{-\frac{1}{2}\mathrm{i}\omega_0 t}\sum_n c_n\left[\int_{-\infty}^{\infty}\psi_n(x)\exp(-\mathrm{i}2\pi sx)\,\mathrm{d}x\right]e^{-\mathrm{i}n\omega_0 t}, \tag{51}$$

and that

$$\int_{-\infty}^{\infty} H_n(\alpha x)e^{-\frac{1}{2}(\alpha x)^2}\exp(-\mathrm{i}2\pi s x)\,\mathrm{d}x = (-\mathrm{i})^n\big(\sqrt{2\pi}/\alpha\big)H_n(2\pi s/\alpha)e^{-\frac{1}{2}(2\pi s/\alpha)^2}, \qquad (52)$$

it is seen that $\psi(x,t)$ evolves in time to become its own Fourier transform after every quarter of a period. A straightforward comparison between Eqs.(49) and (52) shows that the linear scaling needed to convert the $x$-coordinate to the spatial frequency is $s = (\alpha^2/2\pi)x$.

**10. Wigner distribution**. In quantum mechanics, the Wigner function (or distribution) is defined for pure as well as mixed states.[14-16] Our brief introduction to the Wigner function and its basic properties in this section will be limited to the case of a point particle in three-dimensional space, whose pure state is specified by the Schrödinger wave-function $\psi(\boldsymbol{r})$. The equations are somewhat simplified if we write $\mathrm{d}\boldsymbol{r}$ for $\mathrm{d}x\mathrm{d}y\mathrm{d}z$ and $\mathrm{d}\boldsymbol{p}$ for $\mathrm{d}p_x\mathrm{d}p_y\mathrm{d}p_z$. The momentum-space wave-function $\phi(\boldsymbol{p})$ is related to $\psi(\boldsymbol{r})$ via Fourier transformation, as follows:

$$\phi(\boldsymbol{p}) = (2\pi\hbar)^{-3/2}\iiint_{-\infty}^{\infty}\psi(\boldsymbol{r})\exp(-\mathrm{i}\,\boldsymbol{p}\cdot\boldsymbol{r}/\hbar)\,\mathrm{d}\boldsymbol{r}. \qquad (53)$$

$$\psi(\boldsymbol{r}) = (2\pi\hbar)^{-3/2}\iiint_{-\infty}^{\infty}\phi(\boldsymbol{p})\exp(\mathrm{i}\,\boldsymbol{p}\cdot\boldsymbol{r}/\hbar)\,\mathrm{d}\boldsymbol{p}. \qquad (54)$$

The following important property of Dirac's $\delta$-function in spaces of three dimensions is all that is needed to prove some of the fundamental properties of the Wigner function:

$$\iiint_{-\infty}^{\infty}\exp(\pm\,\mathrm{i}2\boldsymbol{p}\cdot\boldsymbol{r}/\hbar)\,\mathrm{d}\boldsymbol{p} = (2\pi)^3\delta(2x/\hbar)\delta(2y/\hbar)\delta(2z/\hbar) = (\pi\hbar)^3\delta(x)\delta(y)\delta(z). \qquad (55)$$

The original motivation for introducing the Wigner distribution was to study quantum corrections to classical statistical mechanics by relating Schrödinger's wave-function to a probability distribution in the $(\boldsymbol{r},\boldsymbol{p})$ phase space. For a pure state with the wave-function $\psi(\boldsymbol{r})$, the Wigner function is defined as follows:

$$W(\boldsymbol{r},\boldsymbol{p}) = (\pi\hbar)^{-3}\iiint_{-\infty}^{\infty}\psi(\boldsymbol{r}-\boldsymbol{r}')\psi^*(\boldsymbol{r}+\boldsymbol{r}')\exp(\mathrm{i}2\boldsymbol{p}\cdot\boldsymbol{r}'/\hbar)\,\mathrm{d}\boldsymbol{r}'. \qquad (56)$$

By virtue of the fact that $W(\boldsymbol{r},\boldsymbol{p})$ is its own complex conjugate, the Wigner function is seen to be real-valued. Also, as is generally the case with conventional probability distributions, the integral of $W(\boldsymbol{r},\boldsymbol{p})$ over the six-dimensional position-momentum space equals 1.0. This is easily demonstrated by invoking Eq.(55) and recalling the sifting property of the $\delta$-function, as follows:

$$\int_{-\infty}^{\infty}W(\boldsymbol{r},\boldsymbol{p})\mathrm{d}\boldsymbol{r}\mathrm{d}\boldsymbol{p} = \int_{-\infty}^{\infty}\psi(\boldsymbol{r}-\boldsymbol{r}')\psi^*(\boldsymbol{r}+\boldsymbol{r}')\delta(\boldsymbol{r}')\mathrm{d}\boldsymbol{r}'\mathrm{d}\boldsymbol{r} = \int_{-\infty}^{\infty}|\psi(\boldsymbol{r})|^2\mathrm{d}\boldsymbol{r} = 1. \qquad (57)$$

In contrast to conventional probability distributions, there exist situations in which $W(\boldsymbol{r},\boldsymbol{p})$ acquires negative values in some regions of the phase space, hence the reason why it is referred to as the Wigner "quasi-probability" distribution. A direct application of the Cauchy-Schwarz inequality[23] to Eq.(56) reveals the confinement of the function to the following interval:

$$-(\pi\hbar)^{-3} \le W(\boldsymbol{r},\boldsymbol{p}) \le (\pi\hbar)^{-3}. \qquad (58)$$

It is possible to express the Wigner function in terms of the momentum-space wave-function $\phi(\boldsymbol{p})$ by substituting $\psi(\boldsymbol{r})$ from Eq.(54) into Eq.(56), rearranging the order of integration, invoking Eq.(55), and using the sifting property of the $\delta$-function. We will find

$$W(\boldsymbol{r},\boldsymbol{p}) = (\pi\hbar)^{-3}\iiint_{-\infty}^{\infty}\phi(\boldsymbol{p}-\boldsymbol{p}')\phi^*(\boldsymbol{p}+\boldsymbol{p}')\exp(-\mathrm{i}2\boldsymbol{p}'\cdot\boldsymbol{r}/\hbar)\,\mathrm{d}\boldsymbol{p}'. \qquad (59)$$

An important characteristic of $W(\boldsymbol{r},\boldsymbol{p})$ is that its integral over $\boldsymbol{p}$ yields the marginal probability density function $|\psi(\boldsymbol{r})|^2$ for the position $\boldsymbol{r}$, whereas its integral over $\boldsymbol{r}$ yields the marginal probability density function $|\phi(\boldsymbol{p})|^2$ for the momentum $\boldsymbol{p}$. Both statements are easily proven, once again, by invoking the identity in Eq.(55) and the sifting property of the $\delta$-function in conjunction with the integral of Eq.(56) over $\boldsymbol{p}$, or that of Eq.(59) over $\boldsymbol{r}$. We find

$$\int_{-\infty}^{\infty} W(\boldsymbol{r},\boldsymbol{p})\mathrm{d}\boldsymbol{p} = (\pi\hbar)^{-3} \iiint_{-\infty}^{\infty} \psi(\boldsymbol{r}-\boldsymbol{r}')\psi^*(\boldsymbol{r}+\boldsymbol{r}')(\pi\hbar)^3\delta(\boldsymbol{r}')\mathrm{d}\boldsymbol{r}' = |\psi(\boldsymbol{r})|^2. \tag{60}$$

$$\int_{-\infty}^{\infty} W(\boldsymbol{r},\boldsymbol{p})\mathrm{d}\boldsymbol{r} = (\pi\hbar)^{-3} \iiint_{-\infty}^{\infty} \phi(\boldsymbol{p}-\boldsymbol{p}')\phi^*(\boldsymbol{p}+\boldsymbol{p}')(\pi\hbar)^3\delta(\boldsymbol{p}')\mathrm{d}\boldsymbol{p}' = |\phi(\boldsymbol{p})|^2. \tag{61}$$

One could arrive at the same results by integrating Eq.(56) over $\boldsymbol{r}$, or Eq.(59) over $\boldsymbol{p}$, after applying a 45° coordinate-system rotation. For instance, in the case of integrating Eq.(56) over $\boldsymbol{r}$, upon introducing the rotated coordinates $\boldsymbol{\rho} = (\boldsymbol{r}-\boldsymbol{r}')/\sqrt{2}$ and $\boldsymbol{\rho}' = (\boldsymbol{r}+\boldsymbol{r}')/\sqrt{2}$, one obtains

$$\begin{aligned}
\int_{-\infty}^{\infty} W(\boldsymbol{r},\boldsymbol{p})\mathrm{d}\boldsymbol{r} &= (\pi\hbar)^{-3} \int_{-\infty}^{\infty} \psi(\boldsymbol{r}-\boldsymbol{r}')\psi^*(\boldsymbol{r}+\boldsymbol{r}') \exp(\mathrm{i}2\boldsymbol{p}\cdot\boldsymbol{r}'/\hbar)\, \mathrm{d}\boldsymbol{r}\mathrm{d}\boldsymbol{r}' \\
&= (\pi\hbar)^{-3} \int_{-\infty}^{\infty} \psi(\sqrt{2}\boldsymbol{\rho})\psi^*(\sqrt{2}\boldsymbol{\rho}') \exp\left[\mathrm{i}\sqrt{2}\boldsymbol{p}\cdot(\boldsymbol{\rho}'-\boldsymbol{\rho})/\hbar\right] \mathrm{d}\boldsymbol{\rho}\mathrm{d}\boldsymbol{\rho}' \\
&= (2\pi\hbar)^{-3}\left[\int_{-\infty}^{\infty} \psi(\boldsymbol{\rho}) \exp(-\mathrm{i}\boldsymbol{p}\cdot\boldsymbol{\rho}/\hbar)\, \mathrm{d}\boldsymbol{\rho}\right] \times \left[\int_{-\infty}^{\infty} \psi^*(\boldsymbol{\rho}') \exp(\mathrm{i}\boldsymbol{p}\cdot\boldsymbol{\rho}'/\hbar)\, \mathrm{d}\boldsymbol{\rho}'\right] \\
&= \phi(\boldsymbol{p})\phi^*(\boldsymbol{p}) = |\phi(\boldsymbol{p})|^2.
\end{aligned} \tag{62}$$

Aside from statistical physics, quantum mechanics, and quantum optics, the Wigner function has found recent applications in other branches of scientific inquiry such as classical optics[17] and electrical, optical, and acoustic signal processing.[18]

---

**Example 1**. Let the Schrödinger wave-function of a point particle in free space be specified as $\psi(\boldsymbol{r}) = \psi_0 \exp[-(z/\zeta_0)^2 + \mathrm{i}k_0 z]$, where $\zeta_0 > 0$ and $k_0$ are real constants. The corresponding Wigner distribution, straightforwardly computed from Eq.(56), is found to be

$$W(\boldsymbol{r},\boldsymbol{p}) = \frac{\zeta_0|\psi_0|^2}{\hbar\sqrt{2\pi}} e^{-2(z/\zeta_0)^2} e^{-[\zeta_0(p_z-\hbar k_0)/\sqrt{2}\hbar]^2} \delta(p_x)\delta(p_y). \tag{63}$$

The uniformity of the above $W(\boldsymbol{r},\boldsymbol{p})$ in the $xy$-plane reflects the uniformity of the particle's probability amplitude in this plane, while the concentration of $W(\boldsymbol{r},\boldsymbol{p})$ around $(p_x, p_y) = (0,0)$ reveals the certainty in these values of the $x$ and $y$ components of momentum in conjunction with the total uncertainty in the particle's position in the $xy$-plane. Along the $z$-axis, the expected location of the particle is the vicinity of $z = 0$ to within a distance of the order of $\pm\zeta_0$, which is manifest in the $z$-dependence of $W(\boldsymbol{r},\boldsymbol{p})$. Similarly, the dependence of the Wigner function on $p_z$ shows the expected value of the $z$-component of the particle's momentum to be $\hbar k_0$, with an uncertainty of the order of $\pm\hbar/\zeta_0$, consistent with the Heisenberg uncertainty principle.[12] It should be emphasized that, although the Wigner function in the present example is non-negative, there exist other situations in which the function acquires negative local values.

---

With proper adjustments for the context, Eqs.(53), (54), (56), and (59) prove useful in a variety of applications such as light propagation in dispersive media, diffraction in the paraxial regime, and studies involving the coherence properties of light.[17,18] As an elementary example, let $f(t)$ be a time-dependent signal whose Fourier transform is given by

$$F(s) = \int_{-\infty}^{\infty} f(t)e^{\mathrm{i}2\pi st}\mathrm{d}t. \tag{64}$$

For the sake of simplicity, $f(t)$ is assumed to be normalized, so that $\int_{-\infty}^{\infty}|f(t)|^2\mathrm{d}t = 1$; Parseval's theorem[1] then ensures that $\int_{-\infty}^{\infty}|F(s)|^2\mathrm{d}s = 1$. The earlier results of the present section remain equally applicable to $f(t)$ and $F(s)$ if we reduce the dimensionality of the functions (and of the integrals) from 3 to 1, replace $\psi(\boldsymbol{r})$ with $f(t)$, substitute $F(s)$ for $\phi(\boldsymbol{p})$, and formally set $\hbar = -1/2\pi$. In accordance with Eq.(56), the Wigner distribution of $f(t)$ will be

$$W_f(t,s) = 2\int_{-\infty}^{\infty} f(t-t')f^*(t+t')e^{-\mathrm{i}4\pi st'}\mathrm{d}t'. \tag{65}$$

Certain operations performed on $f(t)$ result in fairly simple changes in $W_f(t,s)$, so that a knowledge of the Wigner function would enable one to predict the outcome of the operation. For instance, a translation in time, defined as $g(t) = f(t-t_0)$, results in $W_g(t,s) = W_f(t-t_0,s)$. Similarly, multiplication by a linear phase-factor, say, $g(t) = e^{\mathrm{i}2\pi s_0 t}f(t)$, where $s_0$ is a real constant, results in $W_g(t,s) = W_f(t,s+s_0)$. The following examples show some intriguing applications of these ideas in the field of classical optics.

---

**Example 2**. Let $f_0(t)$ be an optical wave-packet in a transparent dispersive medium of refractive index $n(\sigma)$, where the angular frequency of the EM field is taken to be $\omega = 2\pi\sigma$. We denote the propagation $k$-vector along the $z$-axis by $\boldsymbol{k} = \omega n(\omega)\hat{\boldsymbol{z}}/c$, assume that $f_0(t)$ has a reasonably narrow frequency spectrum centered at $\omega_0 = 2\pi\sigma_0$, and that the frequency-dependence of the magnitude of $\boldsymbol{k}$ is specified as $k(\sigma) = k_0 + k_1(\sigma-\sigma_0) + ½k_2(\sigma-\sigma_0)^2$. Upon traveling a distance $z_0$ through the medium, the wave-packet acquires a new functional form, namely,

$$\begin{aligned} f(t; z=z_0) &= 2\mathrm{Re}\int_0^{\infty} F_0(\sigma)e^{\mathrm{i}k(\sigma)z_0}e^{-\mathrm{i}2\pi\sigma t}\mathrm{d}\sigma \\ &= 2\mathrm{Re}\left\{e^{\mathrm{i}(k_0 z_0-\omega_0 t)}\int_0^{\infty} F_0(\sigma)e^{\mathrm{i}[k_1(\sigma-\sigma_0)+½k_2(\sigma-\sigma_0)^2]z_0}e^{-\mathrm{i}2\pi(\sigma-\sigma_0)t}\mathrm{d}\sigma\right\} \\ &= 2\mathrm{Re}\left\{e^{\mathrm{i}(k_0 z_0-\omega_0 t)}\int_{-\infty}^{\infty} \mathcal{E}_0(s)e^{\mathrm{i}(k_1 s+½k_2 s^2)z_0}e^{-\mathrm{i}2\pi st}\mathrm{d}s\right\}. \end{aligned} \tag{66}$$

Here, we have defined the Fourier transform of the envelope $e(t; z=0)$ of the wave-packet as $\mathcal{E}_0(s) = F(\sigma_0+s)$ for $s > -\sigma_0$, then extended the lower limit of the integral to $-\infty$. In other words, $\mathcal{E}_0(s)$ is the positive-frequency half of $F(\sigma)$ whose center, originally at $\sigma = \sigma_0$, has been down-shifted to $\sigma = 0$, followed by a renaming of $\sigma$ to $s$. A glance at Eq.(59) reveals that the multiplication of $\mathcal{E}_0(s)$ by the phase-factor $\exp[\mathrm{i}(k_1 s + ½k_2 s^2)z_0]$ should modify the envelope's Wigner function in accordance with the following formula:

$$W_e(t,s;\ z=z_0) = W_e[t-(k_1 z_0/2\pi)-(k_2 z_0/2\pi)s, s;\ z=0]. \tag{67}$$

The shift along the time coordinate reveals the group velocity of the packet as $v_g = 2\pi/k_1$, while the coordinate transformation via a tilt of the $t$-axis accounts for the wave-packet's distortion caused by the group velocity dispersion. The marginal distribution along the new $t$-axis would be the envelope's intensity profile versus the time $t$ at $z = z_0$. When computing the marginal distribution by integrating along the direction perpendicular to the new (rotated) $t$-axis, one must remember to multiply the end result by the cosine of the angle between the old and the new $t$-axis, as the integration path is now rotated by the same angle away from the $s$-axis.

---

**Example 3**. The scalar light amplitude distribution $a_0(x,y)$ in the $xy$-plane at $z = 0$ with the vacuum wavelength $\lambda_0$ was examined in Sec.3, where its spatial Fourier transform was given by

$$A_0(\sigma_x/\lambda_0, \sigma_y/\lambda_0) = \iint_{-\infty}^{\infty} a_0(x,y)e^{-\mathrm{i}2\pi(\sigma_x x+\sigma_y y)/\lambda_0}\mathrm{d}x\mathrm{d}y. \tag{68}$$

Assuming that $a_0(x,y)$ is properly normalized so that $\iint_{-\infty}^{\infty}|a_0(x,y)|^2\mathrm{d}x\mathrm{d}y = 1$, a comparison with Eq.(52) shows $a_0(x,y)$ to be the two-dimensional counterpart of $\psi(\boldsymbol{r})$, and if we formally set $\hbar = \lambda_0/2\pi$, then $\lambda_0^{-1}A_0(\sigma_x/\lambda_0, \sigma_y/\lambda_0)$ would be the counterpart of $\phi(\boldsymbol{p})$. Thus, in accordance with Eq.(56), the Wigner function of $a_0(x,y)$ could be written as

$$W_{a_0}(x,y;\sigma_x,\sigma_y) = (2/\lambda_0)^2 \iint_{-\infty}^{\infty} a_0(x-x',y-y')a_0^*(x+x',y+y')e^{\mathrm{i}4\pi(\sigma_x x'+\sigma_y y')/\lambda_0}\mathrm{d}x'\mathrm{d}y'. \tag{69}$$

Upon propagation along the $z$-axis, the light amplitude distribution in the $xy$-plane at $z$ will be given by Eq.(7). In the paraxial regime, where $\sigma_z = (1-\sigma_x^2-\sigma_y^2)^{½} \cong 1 - ½(\sigma_x^2+\sigma_y^2)$, we find

$$a(x,y,z_0) \cong \lambda_0^{-2} \iint_{-\infty}^{\infty} A_0(\sigma_x/\lambda_0, \sigma_y/\lambda_0)e^{\mathrm{i}2\pi[1-½(\sigma_x^2+\sigma_y^2)]z_0/\lambda_0}e^{\mathrm{i}2\pi(\sigma_x x+\sigma_y y)/\lambda_0}\mathrm{d}\sigma_x\mathrm{d}\sigma_y. \tag{70}$$

Since, in consequence of propagation, the Fourier transform of $a_0(x,y)$ appearing in Eq.(70) is multiplied by the phase-factor $e^{\mathrm{i}2\pi[1-½(\sigma_x^2+\sigma_y^2)]z_0/\lambda_0}$, the Wigner function of $a(x,y,z_0)$ must be a correspondingly modified version of the Wigner function of $a_0(x,y)$; that is,

$$W_a(x,y;\sigma_x,\sigma_y) \cong W_{a_0}(x-z_0\sigma_x, y-z_0\sigma_y;\sigma_x,\sigma_y). \tag{71}$$

It is seen that a coordinate transformation turns the Wigner function of the initial distribution at $z=0$ into that of the light amplitude profile after propagation by the arbitrary distance $z_0$ — with the caveat that the propagation should take place in the paraxial regime.

---

**11. The van Cittert-Zernike theorem**. This fundamental theorem of classical statistical optics relates the cross-correlation function of the light amplitude distribution in a plane illuminated by a distant, spatially incoherent, quasi-monochromatic source to the Fourier transform of the intensity distribution in the plane of that source.[2,4,22] With reference to Fig.4(a), let the scalar light amplitude distribution in the $xy$-plane at $z=0$ be the real-valued function $a_0(x,y,t)$, where the duration $T$ of the wave-packet emanating from the source could be arbitrarily long. The scaled Fourier transform of the wave-packet at the point $(x,y)$ is given by $A_0(x,y,\omega)$, where

$$a_0(x,y,t) = a(x,y,z=0,t) = \frac{\sqrt{T}}{2\pi}\int_{-\infty}^{\infty} A_0(x,y,\omega)e^{-\mathrm{i}\omega t}\mathrm{d}\omega. \tag{72}$$

Throughout the present section, we take $a_0(x,y,t)$ to be real-valued, which makes its Fourier transform a Hermitian function of the frequency $\omega$; that is, $A_0(x,y,-\omega) = A_0^*(x,y,\omega)$. Invoking Parseval's theorem[1] and introducing the time-averaged intensity distribution $I_0(x,y)$, we now write

$$I_0(x,y) = T^{-1}\int_{-\infty}^{\infty} a_0^2(x,y,t)\mathrm{d}t = (2\pi)^{-1}\int_{-\infty}^{\infty} A_0(x,y,\omega)A_0^*(x,y,\omega)\mathrm{d}\omega. \tag{73}$$

For a quasi-monochromatic source, $A_0(x,y,\omega) = |A_0|\exp(\mathrm{i}\varphi_0)$ has a narrow linewidth $\Delta\omega$ around a central frequency $\pm\omega_0$, as shown in Fig.4(b). For natural light, the phase $\varphi_0(x,y,\omega)$ is a rapidly varying function of $\omega$ that assumes many different (seemingly random) values between 0 and $2\pi$ as one moves along the $\omega$-axis within the bandwidth $\Delta\omega$. If, in addition, the source is spatially incoherent, $\varphi_0(\omega)$ at $(x,y)$ will have no relation to $\varphi_0(\omega)$ at another location $(x',y')$ of the source. All in all, one can state, to a good approximation, that a spatially incoherent natural light source sitting in the $xy$-plane at $z=0$ and radiating a wave-packet of sufficiently long duration $T$ satisfies the identity

$$\int_{-\infty}^{\infty} A_0(x,y,\omega)A_0^*(x',y',\omega)\mathrm{d}\omega = 2\pi I_0(x,y)\delta(x-x')\delta(y-y'). \tag{74}$$

The above equation embodies the fact that, within the bandwidth $\Delta\omega$ of the light source, $\varphi_0(x,y,\omega)$ and $\varphi_0(x',y',\omega)$, considered as functions of $\omega$, vary rapidly and independently of each other when $(x,y) \neq (x',y')$, thus causing the integral over $\omega$ to essentially vanish. However, when $(x,y) = (x',y')$, the phase $\varphi_0$ is eliminated and the integral over $\omega$ becomes $2\pi I_0(x,y)$, which is what one obtains by integrating the right-hand side of Eq.(74) over $x'$ and $y'$.

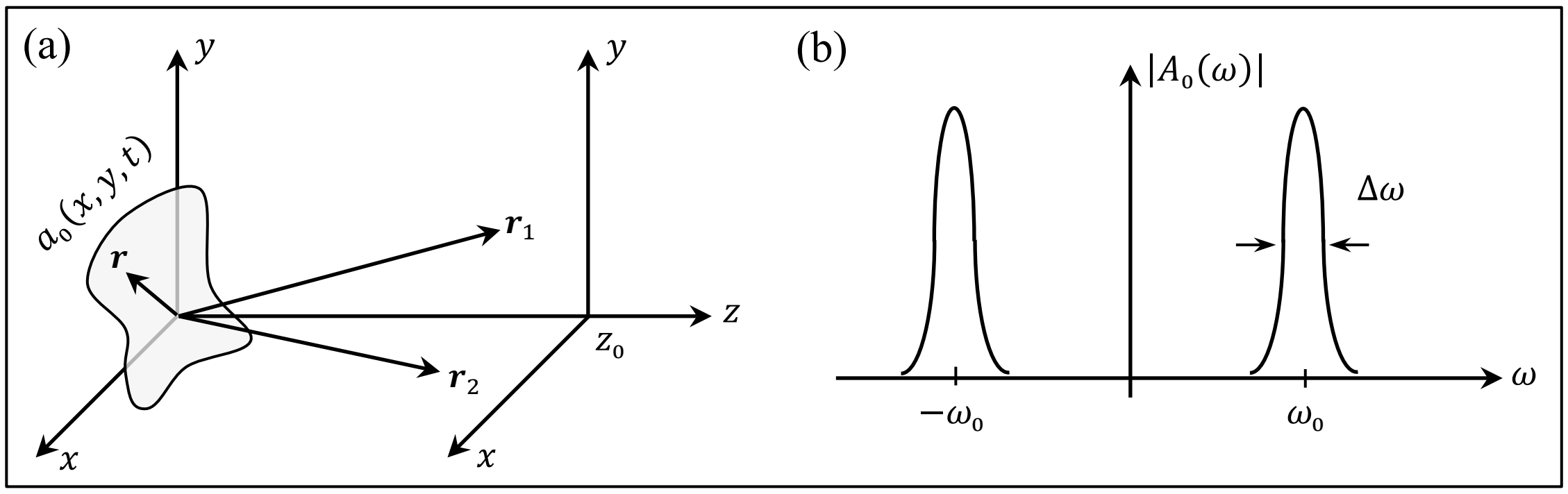

**Fig.4**. (a) A quasi-monochromatic, spatially incoherent, planar light source sits in the $xy$-plane at $z = 0$. The scalar light amplitude in the source plane is $a_0(x,y,t)$, and the cross-correlation function between the amplitudes at $\boldsymbol{r}_1$ and $\boldsymbol{r}_2$, a pair of points in a distant $xy$-plane at $z = z_0$, is denoted by $\Gamma(\boldsymbol{r}_1, \boldsymbol{r}_2, \tau)$. (b) The spectral density of the quasi-monochromatic light at all points $(x,y,z=0)$ of the source is a narrow-band function of the frequency $\omega$, centered at $\omega = \pm\omega_0$ and having a bandwidth $\Delta\omega \ll \omega_0$.

Next, let us consider the optical field amplitude profile in the $xy$-plane at a large distance $z_0$ from the source plane. Figure 4(a) shows the points $\boldsymbol{r} = (x,y,0)$ in the source plane and $\boldsymbol{r}_1 = (x_1, y_1, z_0)$ in the destination plane, with the distance between them being approximated as

$$|\boldsymbol{r}_1 - \boldsymbol{r}| = \sqrt{(x_1 - x)^2 + (y_1 - y)^2 + z_0^2} \cong z_0 + [(x_1 - x)^2 + (y_1 - y)^2]/2z_0. \qquad (75)$$

The field amplitude at $\boldsymbol{r}_1$ is obtained by a superposition of all the rays that emanate from various locations on the source and arrive at $\boldsymbol{r}_1$ after propagating in free space. We thus have

$$a(x_1, y_1, z_0, t) = \frac{\sqrt{T}}{2\pi} \iiint_{-\infty}^{\infty} |\boldsymbol{r}_1 - \boldsymbol{r}|^{-1} A_0(x,y,\omega) \exp[\mathrm{i}(\omega/c)(|\boldsymbol{r}_1 - \boldsymbol{r}| - ct)]\, \mathrm{d}x\mathrm{d}y\mathrm{d}\omega$$
$$\cong \frac{\sqrt{T}}{2\pi z_0} \iiint_{-\infty}^{\infty} A_0(x,y,\omega) e^{\mathrm{i}\omega z_0/c} e^{\mathrm{i}\omega[(x_1-x)^2+(y_1-y)^2]/2z_0c} e^{-\mathrm{i}\omega t} \mathrm{d}x\mathrm{d}y\mathrm{d}\omega. \qquad (76)$$

The cross-correlation function between the field amplitudes at points $\boldsymbol{r}_1$ and $\boldsymbol{r}_2$ in the destination plane (including a time delay $\tau = t_2 - t_1$) is now given by

$$\Gamma(x_1, y_1; x_2, y_2; \tau) = T^{-1} \int_{-\infty}^{\infty} a(x_1, y_1, z_0, t) a(x_2, y_2, z_0, t + \tau) \mathrm{d}t$$
$$\cong (2\pi z_0)^{-2} \int_{-\infty}^{\infty} A_0(x,y,\omega) A_0(x',y',\omega') e^{\mathrm{i}(\omega+\omega')z_0/c}$$
$$\times e^{\mathrm{i}\omega[(x_1-x)^2+(y_1-y)^2]/2z_0c} \times e^{\mathrm{i}\omega'[(x_2-x')^2+(y_2-y')^2]/2z_0c}$$
$$\times e^{-\mathrm{i}\omega'\tau} \left[\int_{-\infty}^{\infty} e^{-\mathrm{i}(\omega+\omega')t} \mathrm{d}t\right] \mathrm{d}x'\mathrm{d}y'\mathrm{d}\omega'\mathrm{d}x\mathrm{d}y\mathrm{d}\omega. \qquad (77)$$

Given that $\int_{-\infty}^{\infty} e^{-\mathrm{i}(\omega+\omega')t} \mathrm{d}t = 2\pi\delta(\omega + \omega')$, carrying out the integral over $\omega'$ in Eq.(77) leads to

$$\Gamma(x_1, y_1; x_2, y_2; \tau) \cong (2\pi z_0^2)^{-1} \int_{-\infty}^{\infty} A_0(x,y,\omega) A_0^*(x',y',\omega)$$
$$\times e^{\mathrm{i}\omega[(x_1^2+y_1^2) - (x_2^2+y_2^2)]/2z_0c} \times e^{\mathrm{i}\omega[(x^2+y^2) - (x'^2+y'^2)]/2z_0c}$$
$$\times e^{-\mathrm{i}\omega(xx_1+yy_1-x'x_2-y'y_2)/z_0c} \times e^{\mathrm{i}\omega\tau} \mathrm{d}x'\mathrm{d}y'\mathrm{d}x\mathrm{d}y\mathrm{d}\omega. \qquad (78)$$

The exponential factors appearing in the integrand of Eq.(78) do not change very much over the narrow bandwidth $\Delta\omega$ of the source, thus allowing one to set $\omega = \pm\omega_0$ in all four exponential factors. (The implicit assumption here is that $\tau\Delta\omega \ll 1$, and that $z_0$ is sufficiently large compared to the source diameter and also compared to $x_1$, $y_1$, $x_2$, and $y_2$.) Substitution from Eq.(74) into Eq.(78) followed by integration over $x'$ and $y'$ now yields

$$\Gamma(x_1, y_1; x_2, y_2; \tau) \cong z_0^{-2}\mathrm{Re}\big\{e^{\mathrm{i}\omega_0[(x_1^2+y_1^2)\,-\,(x_2^2+y_2^2)]/2z_0c} \times e^{\mathrm{i}\omega_0\tau}$$
$$\times \iint_{-\infty}^{\infty} I_0(x, y)e^{\mathrm{i}\omega_0[(x_2-x_1)x\,+\,(y_2-y_1)y]/z_0c}\mathrm{d}x\mathrm{d}y\big\}. \qquad (79)$$

To streamline this fundamental result of the van Cittert-Zernike theorem, we define the normalized source intensity $\hat{I}_0(x, y)$, the time-averaged intensities at $\boldsymbol{r}_1$ and $\boldsymbol{r}_2$, the auxiliary phase term $\psi_{12}$, and the complex degree of mutual coherence $\gamma_{12}$, as follows:

$$\hat{I}_0(x, y) = I_0(x, y)/\iint_{-\infty}^{\infty} I_0(x, y)\mathrm{d}x\mathrm{d}y. \qquad (80a)$$

$$I(\boldsymbol{r}_1) = \Gamma(x_1, y_1; x_1, y_1; 0) = z_0^{-2} \iint_{-\infty}^{\infty} I_0(x, y)\mathrm{d}x\mathrm{d}y. \qquad (80b)$$

$$I(\boldsymbol{r}_2) = \Gamma(x_2, y_2; x_2, y_2; 0) = z_0^{-2} \iint_{-\infty}^{\infty} I_0(x, y)\mathrm{d}x\mathrm{d}y. \qquad (80c)$$

$$\psi_{12}(x_1, y_1; x_2, y_2) = \pi[(x_1^2 + y_1^2) - (x_2^2 + y_2^2)]/\lambda_0 z_0. \qquad (80d)$$

$$\gamma_{12}(x_1, y_1; x_2, y_2) = \iint_{-\infty}^{\infty} \hat{I}_0(x, y) \exp\{\mathrm{i}2\pi[(x_2 - x_1)x + (y_2 - y_1)y]/(\lambda_0 z_0)\}\, \mathrm{d}x\mathrm{d}y. \qquad (80e)$$

In these equations, $\lambda_0 = 2\pi c/\omega_0$ is the vacuum wavelength of the source at its center frequency $\omega_0$. It is seen that the complex degree of mutual coherence $\gamma_{12}$ is the $2D$ Fourier transform of the normalized source intensity $\hat{I}_0(x, y)$ evaluated at the spatial frequency $(s_x, s_y) = [(x_2 - x_1)/(\lambda_0 z_0), (y_2 - y_1)/(\lambda_0 z_0)]$. The normalized correlation function may now be written as

$$\Gamma(\boldsymbol{r}_1, \boldsymbol{r}_2, \tau)/I(\boldsymbol{r}_1) = \mathrm{Re}(\gamma_{12}e^{\mathrm{i}\psi_{12}}e^{\mathrm{i}\omega_0\tau}). \qquad (81)$$

If one installs an opaque screen in the $xy$-plane at $z = z_0$ with two pinhole apertures at $\boldsymbol{r}_1$ and $\boldsymbol{r}_2$, as depicted in Fig.5, an interference pattern will form in the $xy$-plane at a reasonably long distance $z_1$ behind the screen. Denoting by $d_{12}$ the separation between the pinholes, by $\ell_1, \ell_2$ the distances from the pinholes to the observation point $\zeta$ within the fringe pattern, and by $\theta$ the small angle subtended by $\zeta$ at the midpoint of the pinholes, it is easy to see that the time delay associated with the two rays that emerge from the pinholes and arrive at $\zeta$ is $\tau = (\ell_1 - \ell_2)/c \cong d_{12}\theta/c$. The fringe intensity at $\zeta$, expressed as a function of the angular position $\theta$, is thus proportional to

$$I(\theta) \cong T^{-1}\int_{-\infty}^{\infty}[a(\boldsymbol{r}_1,t) + a(\boldsymbol{r}_2,t+\tau)]^2\mathrm{d}t = I(\boldsymbol{r}_1) + I(\boldsymbol{r}_2) + 2\Gamma(\boldsymbol{r}_1,\boldsymbol{r}_2,\tau)$$

$$= 2I(\boldsymbol{r}_1)\{1 + |\gamma_{12}|\cos[(2\pi d_{12}/\lambda_0)\theta + \phi_{\gamma_{12}} + \psi_{12}]\}. \quad (82)$$

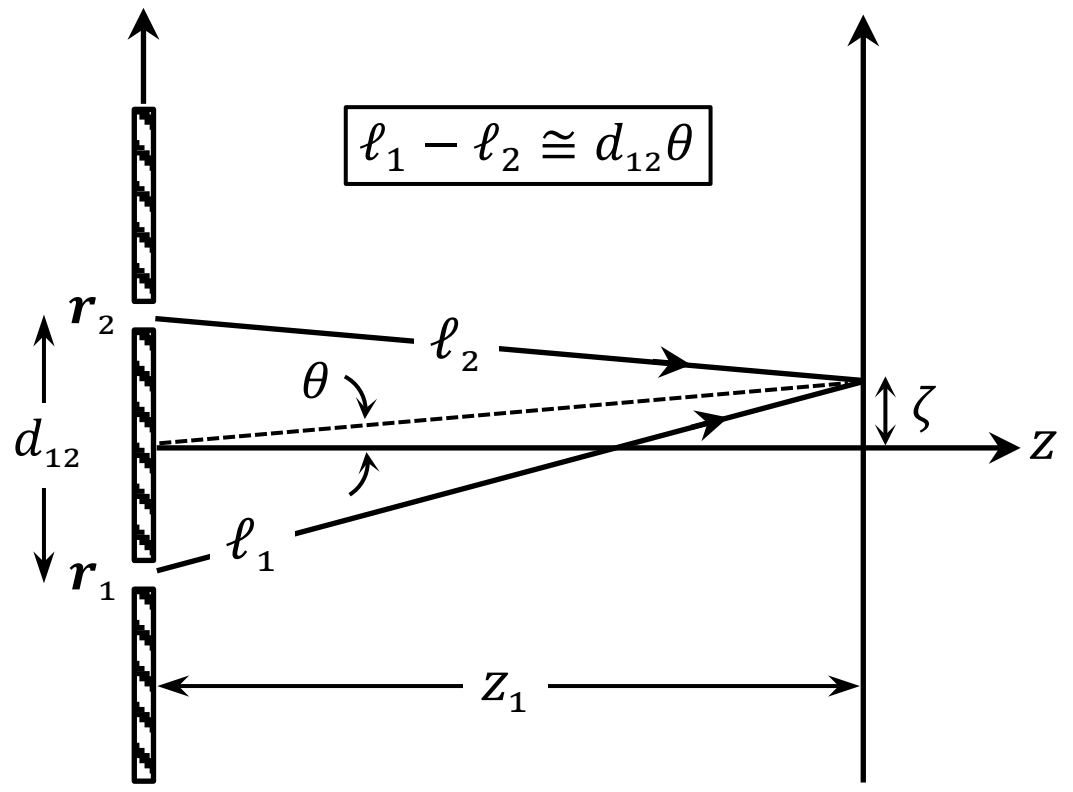

**Fig.5**. A pair of pinholes at $\boldsymbol{r}_1 = (x_1, y_1, z_0)$ and $\boldsymbol{r}_2 = (x_2, y_2, z_0)$ produces interference fringes in the $xy$-plane at a reasonably large distance $z_1$ along the $z$-axis. The time delay due to propagation from the pinholes to the observation point $\zeta$ is $\tau = (\ell_1 - \ell_2)/c \cong d_{12}\theta/c$, where $d_{12}$ is the distance between the pinholes, $\theta$ is the angle subtended by $\zeta$ at the midpoint of the pinholes, and $c$ is the speed of light in vacuum.

It is seen that the fringe visibility equals $|\gamma_{12}|$, the absolute value of the complex degree of mutual coherence, which is defined in Eq.(80e) as the $2D$ Fourier transform of the normalized intensity distribution $\hat{I}_0(x,y)$ of our incoherent, quasi-monochromatic light source.

**12. Rate of decline of a bandlimited $\boldsymbol{f(x)}$ when $|\boldsymbol{x}| \to \infty$.** A bandlimited $f(x)$ is a function whose Fourier spectrum $F(s)$ identically vanishes for $|s| > s_0$. It is well known that such functions cannot be confined to a finite interval on the $x$-axis, and that their tails extend all the way to infinity.[1] An elementary example is provided by $\mathrm{sinc}(x) = \sin(\pi x)/(\pi x)$, whose Fourier transform $\mathrm{rect}(s)$ is zero for $|s| > ½$. As $|x| \to \infty$, the magnitude of $\mathrm{sinc}(x)$ declines as $1/x$. A closely related function, $f(x) = \mathrm{sinc}^2(x)$, with $F(s) = \mathrm{rect}(s) * \mathrm{rect}(s) = 1 - |s|$ for $|s| \le 1$, and zero elsewhere, drops as $1/x^2$ when $|x| \to \infty$. Raising $\mathrm{sinc}(x)$ to an arbitrarily large power $n$, one finds that $\mathrm{sinc}^n(x)$, whose tails drop as $1/x^n$ when $|x| \to \infty$, is a bandlimited function with a spectrum that smoothly goes to zero at the edges $s = \pm n/2$ of its bandwidth. Other examples include the zeroth-order Bessel function of the first kind, $J_0(2\pi x)$, whose spectrum $(\pi\sqrt{1-s^2}\,)^{-1}$ is confined to $|s| < 1$, and declines as $|x|^{-½}$ in the tails, and $J_1(2\pi x)/2x$, with the spectral function $\sqrt{1-s^2}$ over the interval $|s| \le 1$, which declines as $|x|^{-3/2}$ when $|x| \to \infty$.[1]

Contrast the behavior of these bandlimited functions with that of the Gaussian, $\exp(-\pi x^2)$, whose tails drop extremely rapidly, but of course the Gaussian, whose Fourier transform is $\exp(-\pi s^2)$, is *not* bandlimited. The question arises as to whether bandlimited functions exist that decline as rapidly as the Gaussian and, if not, what would be the fastest possible decay rate for bandlimited functions.

While we have not been able to find an answer to the above question, we did examine the asymptotic behavior of the function $f(x)$ whose Fourier spectrum over its bandwidth $|s| \le s_0$ is $F(s) = \exp[2s_0/(s^2 - s_0^2)]$. This extremely smooth spectrum has the unusual property that all its derivatives vanish as $s$ approaches the edges of the band at $\pm s_0$. In the limit where $|x| \to \infty$, the tails of $f(x)$ were found to decline exponentially rapidly as $|x|^{-¾}e^{-\sqrt{4\pi|x|}}$ (see Appendix D).

**13. Concluding remarks**. In the early 1990s, Professor Ronald Bracewell of Stanford University visited our college and, during his colloquium, mentioned a curious observation he had made about differentiation. Upon differentiating a function such as $f(x)$, one typically obtains the first derivative, the second derivative, the third derivative, and so on. But, Prof. Bracewell asked, "Could one speak of a fractional derivative, say, a half-derivative?" Bracewell's idea was that, if $n^{\text{th}}$ order differentiation of $f(x)$ entails the multiplication of its Fourier transform $F(s)$ by $(\mathrm{i}2\pi s)^n$, then the half-derivative of $f(x)$ could be naturally defined as the inverse Fourier transform of $\sqrt{\mathrm{i}2\pi s}F(s)$. While I do not know of any practical applications of this idea, I have always thought of it as an interesting mathematical concept which may someday find an application. In what follows, in lieu of final remarks, I derive the half-derivatives of Dirac's $\delta$- function, the rectangular pulse function, and the step function, and proceed to comment on one of their curious properties. The half-derivative of the $\delta$-function is

$$\delta^{(½)}(x) = \frac{\mathrm{d}^{½}}{\mathrm{d}x^{½}}\delta(x) = \int_{-\infty}^{\infty}(\mathrm{i}2\pi s)^{½}\exp(\mathrm{i}2\pi sx)\,\mathrm{d}s$$

$$= 2\sqrt{2\pi}\left[e^{-\mathrm{i}\pi/4}\int_0^{\infty}\zeta^2\exp(-\mathrm{i}2\pi x\zeta^2)\,\mathrm{d}\zeta + e^{\mathrm{i}\pi/4}\int_0^{\infty}\zeta^2\exp(\mathrm{i}2\pi x\zeta^2)\,\mathrm{d}\zeta\right]$$

$$= -4\sqrt{2\pi}\,\mathrm{step}(x)\int_0^{\infty}r^2\exp(-2\pi xr^2)\,\mathrm{d}r$$

$$= -\sqrt{2/\pi}\ \mathrm{step}(x)\,x^{-1}\int_0^{\infty}\exp(-2\pi xr^2)\,\mathrm{d}r = -\frac{\mathrm{step}(x)}{\sqrt{4\pi x^3}}. \tag{83}$$

In going from the first to the second line of the above equation, we have changed the variable to $\zeta = \sqrt{-s}$ for negative $s$, and to $\zeta = \sqrt{s}$ for positive $s$. The integrals in the second line are evaluated on $\pm 45°$ lines in the complex plane, where $\zeta = re^{\pm\mathrm{i}\pi/4}$. The two integrals cancel out when $x < 0$, hence the appearance of the step-function $\mathrm{step}(x)$ on the third line. The integral on the third line is evaluated using the method of integration by parts.

The function $\mathrm{rect}(x)$ is defined as 1 for $|x| < ½$ and 0 elsewhere. Its Fourier transform is readily found to be $\mathrm{sinc}(s) = \sin(\pi s)/(\pi s)$, and its half-derivative, computed in Appendix E, is

$$\frac{\mathrm{d}^{½}}{\mathrm{d}x^{½}}\mathrm{rect}(x) = \int_{-\infty}^{\infty}(\mathrm{i}2\pi s)^{½}\,\mathrm{sinc}(s)\exp(\mathrm{i}2\pi sx)\,\mathrm{d}s = \frac{\mathrm{step}(x+½)}{\sqrt{\pi(x+½)}} - \frac{\mathrm{step}(x-½)}{\sqrt{\pi(x-½)}}. \tag{84}$$

Considering that $\mathrm{rect}(x) = \mathrm{step}(x+½) - \mathrm{step}(x-½)$, the above result indicates that the half-derivative of $\mathrm{step}(x)$ is $\mathrm{step}(x)/\sqrt{\pi x}$. Note that differentiating $\mathrm{step}(x)$ yields $\delta(x)$, and that the derivative of $\mathrm{step}(x)/\sqrt{\pi x}$ is $-\mathrm{step}(x)/\sqrt{4\pi x^3}$, which is the half-derivative of $\delta(x)$.

It is tempting to state that, since $\delta^{(½)}(x) = -\mathrm{step}(x)/\sqrt{4\pi x^3}$ is the inverse transform of $\sqrt{\mathrm{i}2\pi s}$, the half-derivative of any function is its convolution with $\delta^{(½)}(x)$. This interpretation, however, is not quite correct due to the strong singularity of $\delta^{(½)}(x)$ at $x = 0$. As an example, consider the function $\mathrm{step}(x)$, whose convolution with $\delta^{(½)}(x)$ is zero for $x < 0$, as expected, but for $x > 0$ it becomes

$$\mathrm{step}(x) * \delta^{(½)}(x) = \int_0^x \delta^{(½)}(y)\mathrm{d}y = -\frac{1}{\sqrt{4\pi}}\int_0^x y^{-3/2}\mathrm{d}y = \frac{y^{-½}}{\sqrt{\pi}}\bigg|_0^x = \frac{1}{\sqrt{\pi x}} - \infty. \tag{85}$$

The appearance of the unwanted $-\infty$ on the right-hand side of the above equation is what makes the idea of computing $f^{(½)}(x)$ by convolving $f(x)$ with $\delta^{(½)}(x)$ somewhat problematic.

**Acknowledgement**. The author is grateful to Per Jakobsen for insightful observations and for bringing Ref.[24] to his attention; to Brian Anderson for pointing out the example used in Sec.9; and to Sir Michael Berry for deriving the asymptotic form of the bandlimited function described in Appendix D.

## References

1. R. N. Bracewell, *The Fourier Transform and Its Applications* (3rd edition), McGraw-Hill, New York (1999).
2. L. Mandel and E. Wolf, *Optical Coherence and Quantum Optics*, Cambridge University Press, Cambridge, U.K. (1995).
3. J. W. Goodman, *Introduction to Fourier Optics* (4th edition), W. H. Freeman and Company, New York (2017).
4. M. Born and E. Wolf, *Principles of Optics* (7th edition), Cambridge University Press, Cambridge, U.K. (2002).
5. J. D. Jackson, *Classical Electrodynamics* (third edition), Wiley, New York (1999).
6. M. Mansuripur, *Field, Force, Energy and Momentum in Classical Electrodynamics* (revised edition), Bentham Science Publishers, Sharjah, UAE (2017).
7. R. P. Feynman, R. B. Leighton, and M. Sands, *The Feynman Lectures on Physics*, Volume II, Addison-Wesley, Reading, Massachusetts (1964).
8. W. C. Chew, *Waves and Fields in Inhomogeneous Media*, IEEE Press, Piscataway, New Jersey (1995).
9. P. C. Clemmow, *The Plane Wave Spectrum Representation of Electromagnetic Fields*, IEEE Press, Piscataway, New Jersey (1996).
10. T. B. Hansen and A. D. Yaghjian, *Plane-Wave Theory of Time-Domain Fields: Near-Field Scanning Applications*, IEEE Press, Piscataway, New Jersey (1999).
11. R. F. Harrington, *Time-Harmonic Electromagnetic Fields*, IEEE Press, Piscataway, New Jersey (2001).
12. R. L. White, *Basic Quantum Mechanics*, McGraw-Hill, New York (1966).
13. G. Grynberg, A. Aspect, and C. Fabre, *Introduction to Quantum Optics*, Cambridge University Press, Cambridge, U.K. (2010).
14. E. P. Wigner, "On the quantum correction for thermodynamic equilibrium," *Phys. Rev.* **40**, 749–759 (1932).
15. H. Weyl, *The Theory of Groups and Quantum Mechanics*, Dover, New York (1931).
16. J. E. Moyal, "Quantum mechanics as a statistical theory," *Proceedings of the Cambridge Philosophical Society* **45**, 99–124 (1949).
17. M. A. Alonso, "Wigner functions in optics: describing beams as ray bundles and pulses as particle ensembles," *Advances in Optics and Photonics* **3**, 272-365 (2011).
18. W. Mecklenbräuker and F. Hlawatsch, *The Wigner Distribution: Theory and Applications in Signal Processing*, Elsevier, Amsterdam (1997).
19. M. V. Berry, "Faster than Fourier," *Quantum Coherence and Reality: Celebration of the 60th Birthday of Yakir Aharonov*, edited by J. S. Anandan and J. L. Safko (Singapore: World Scientific), pp 55-65 (1994).
20. M. V. Berry and N. I. Zheludev, *Roadmap on Superoscillations* (edited collection of papers), *J. Opt.* **21**, 053002 (2019).
21. M. Mansuripur and P. K. Jakobsen, "An approach to constructing super oscillatory functions," *J. Phys. A. Math. Theor.* **52**, 305202 (2019).
22. M. Mansuripur, *Classical Optics and its Applications* (2nd edition), Cambridge University Press, Cambridge, U.K. (2009).
23. M. Mansuripur, *Mathematical Methods in Science and Engineering: Applications in Optics and Photonics*, Cognella Academic Publishing, San Diego, California (2020).
24. P. P. Vaidyanathan, "Eigenfunctions of the Fourier Transform," *IETE Journal of Education* **49**, 51-58 (2008).
25. F. B. Hildebrand, Advanced Calculus for Applications (2nd edition), Prentice-Hall, Englewood Cliffs, New Jersey (1976).
26. I. S. Gradshteyn and I. M. Ryzhik, *Table of Integrals, Series, and Products* (7th edition), Academic, Amsterdam (2007).

## Appendix A

This appendix provides detailed derivations of some of the stated results in the Example at the end of Sec.7. The electrical current-density $\boldsymbol{J}_{\text{free}}(\boldsymbol{r},t)$ is Fourier transformed as follows:

$$\begin{aligned}\boldsymbol{J}_{\text{free}}(\boldsymbol{k},\omega) &= \int_{-\infty}^{\infty} J_{s0}\hat{\boldsymbol{x}}\cos(k_{y0}y-\omega_0 t)\,\delta(z)\exp[-\mathrm{i}(\boldsymbol{k}\cdot\boldsymbol{r}-\omega t)]\,\mathrm{d}x\mathrm{d}y\mathrm{d}z\mathrm{d}t\\ &= \tfrac{1}{2}J_{s0}\hat{\boldsymbol{x}}\int_{-\infty}^{\infty}[e^{\mathrm{i}(k_{y0}y-\omega_0 t)}+e^{-\mathrm{i}(k_{y0}y-\omega_0 t)}]e^{-\mathrm{i}(k_x x+k_y y-\omega t)}\mathrm{d}x\mathrm{d}y\mathrm{d}t\\ &= \tfrac{1}{2}J_{s0}\hat{\boldsymbol{x}}\left(\int_{-\infty}^{\infty}e^{-\mathrm{i}k_x x}\mathrm{d}x\right)\iint_{-\infty}^{\infty}[e^{-\mathrm{i}(k_y-k_{y0})y}e^{\mathrm{i}(\omega-\omega_0)t}+e^{-\mathrm{i}(k_y+k_{y0})y}e^{\mathrm{i}(\omega+\omega_0)t}]\mathrm{d}y\mathrm{d}t\\ &= 4\pi^3 J_{s0}\hat{\boldsymbol{x}}\,[\delta(k_x)\delta(k_y-k_{y0})\delta(\omega-\omega_0)+\delta(k_x)\delta(k_y+k_{y0})\delta(\omega+\omega_0)].\end{aligned} \quad \text{(A1)}$$

The inverse transformation of the vector potential $\boldsymbol{A}(\boldsymbol{k},\omega)=\mu_0\boldsymbol{J}_{\text{free}}(\boldsymbol{k},\omega)/[k^2-(\omega/c)^2]$ requires two steps. In the first step, the relevant integrals are evaluated and a solution is found that has no contributions from the free-space solutions of Maxwell's equations in the absence of all sources of radiation. The free space solutions (corresponding to $k=\omega/c$) are then added in the second step to arrive at the complete vector potential. The inverse Fourier integral yields

$$\begin{aligned}\boldsymbol{A}(\boldsymbol{r},t) &= (2\pi)^{-4}\int_{-\infty}^{\infty}\boldsymbol{A}(\boldsymbol{k},\omega)\exp[\mathrm{i}(\boldsymbol{k}\cdot\boldsymbol{r}-\omega t)]\,\mathrm{d}k_x\mathrm{d}k_y\mathrm{d}k_z\mathrm{d}\omega\\ &= \frac{\mu_0 J_{s0}\hat{\boldsymbol{x}}}{4\pi}\int_{-\infty}^{\infty}\frac{\delta(k_y-k_{y0})\delta(\omega-\omega_0)+\delta(k_y+k_{y0})\delta(\omega+\omega_0)}{k_y^2+k_z^2-(\omega/c)^2}e^{\mathrm{i}(k_y y+k_z z-\omega t)}\mathrm{d}k_y\mathrm{d}k_z\mathrm{d}\omega\\ &= \left(\frac{\mu_0}{2\pi}\right)J_{s0}\hat{\boldsymbol{x}}\cos(k_{y0}y-\omega_0 t)\int_{-\infty}^{\infty}\frac{\exp(\mathrm{i}k_z z)}{k_z^2+k_{y0}^2-(\omega_0/c)^2}\mathrm{d}k_z.\end{aligned} \quad \text{(A2)}$$

Upon defining $k_{z0}=\sqrt{(\omega_0/c)^2-k_{y0}^2}$ when $|k_{y0}|<\omega_0/c$, and $k_{z0}=\sqrt{k_{y0}^2-(\omega_0/c)^2}$ when $|k_{y0}|>\omega_0/c$, the remaining integral, when evaluated using complex-plane techniques,[23,25] yields

$$\begin{aligned}\boldsymbol{A}(\boldsymbol{r},t) &= \frac{\mu_0 J_{s0}\hat{\boldsymbol{x}}}{2k_{z0}}\begin{cases}-\operatorname{sign}(z)\cos(k_{y0}y-\omega_0 t)\sin(k_{z0}z); & |k_{y0}|<\omega_0/c\\ e^{-k_{z0}|z|}\cos(k_{y0}y-\omega_0 t); & |k_{y0}|>\omega_0/c\end{cases}\\ &= \frac{\mu_0 J_{s0}\hat{\boldsymbol{x}}}{4k_{z0}}\begin{cases}\operatorname{sign}(z)[\sin(k_{y0}y-k_{z0}z-\omega_0 t)-\sin(k_{y0}y+k_{z0}z-\omega_0 t)]; & |k_{y0}|<\omega_0/c\\ 2e^{-k_{z0}|z|}\cos(k_{y0}y-\omega_0 t); & |k_{y0}|>\omega_0/c.\end{cases}\end{aligned} \quad \text{(A3)}$$

For $|k_{y0}|>\omega_0/c$, the vector potential thus obtained is complete and needs no further modifications. In the case of $|k_{y0}|<\omega_0/c$, however, a free-space solution of Maxwell's equations (corresponding to the singularity at $k=\omega/c$) must be added to eliminate the backward propagating wave. Direct inspection of Eq.(A3) shows the required free-space solution to be

$$\boldsymbol{A}(\boldsymbol{r},t) = -\frac{\mu_0 J_{s0}\hat{\boldsymbol{x}}}{4k_{z0}}\big[\sin(k_{y0}y-k_{z0}z-\omega_0 t)+\sin(k_{y0}y+k_{z0}z-\omega_0 t)\big]. \quad \text{(A4)}$$

The complete vector potential is now obtained by adding the free-space solution of Eq.(A4) to that in Eq.(A3), yielding

$$\boldsymbol{A}(\boldsymbol{r},t) = \frac{\mu_0 J_{s0}\hat{\boldsymbol{x}}}{2k_{z0}}\begin{cases}-\sin(k_{y0}y+k_{z0}|z|-\omega_0 t); & |k_{y0}|<\omega_0/c\\ e^{-k_{z0}|z|}\cos(k_{y0}y-\omega_0 t); & |k_{y0}|>\omega_0/c.\end{cases} \quad \text{(A5)}$$

The vector potential thus obtained corresponds, in the case of $|k_{y0}|<\omega_0/c$, to a pair of propagating plane-waves that emanate from the current-carrying sheet into the half-spaces on the

right- and left-hand sides of the sheet. In the case of $|k_{y0}| > \omega_0/c$, the emerging fields are evanescent, residing in the immediate vicinity and on both sides of the current sheet.

## Appendix B

Equation (38) in Sec.8 is derived from Maxwell's 2nd equation after substitutions from the 3rd and 4th equations, as follows:

$$\boldsymbol{k}\times[\boldsymbol{k}\times\boldsymbol{B}(\boldsymbol{k},\omega)] = -\mu_0\omega\boldsymbol{k}\times\boldsymbol{D}(\boldsymbol{k},\omega) + \boldsymbol{k}\times[\boldsymbol{k}\times\boldsymbol{M}(\boldsymbol{k},\omega)]$$

$$\rightarrow [\overbrace{\boldsymbol{k}\cdot\boldsymbol{B}(\boldsymbol{k},\omega)}^{0}]\boldsymbol{k} - k^2\boldsymbol{B}(\boldsymbol{k},\omega) = -\mu_0\varepsilon_0\omega^2\boldsymbol{B}(\boldsymbol{k},\omega) - \mu_0\omega\boldsymbol{k}\times\boldsymbol{P}(\boldsymbol{k},\omega) + \boldsymbol{k}\times[\boldsymbol{k}\times\boldsymbol{M}(\boldsymbol{k},\omega)]$$

$$\rightarrow \quad (k^2 - \omega^2/c^2)\boldsymbol{B}(\boldsymbol{k},\omega) = \mu_0\omega\boldsymbol{k}\times\boldsymbol{P}(\boldsymbol{k},\omega) - \boldsymbol{k}\times[\boldsymbol{k}\times\boldsymbol{M}(\boldsymbol{k},\omega)]$$

$$\rightarrow \quad \boldsymbol{B}(\boldsymbol{k},\omega) = \mu_0\omega\boldsymbol{k}\times\frac{\boldsymbol{P}(\boldsymbol{k},\omega)}{k^2-(\omega/c)^2} - \boldsymbol{k}\times\left[\boldsymbol{k}\times\frac{\boldsymbol{M}(\boldsymbol{k},\omega)}{k^2-(\omega/c)^2}\right]. \tag{B1}$$

Similarly, Eq.(41) is derived from Maxwell's 3rd equation after substitutions from the 1st and 2nd equations, that is,

$$\boldsymbol{k}\times[\boldsymbol{k}\times\boldsymbol{D}(\boldsymbol{k},\omega)] = \varepsilon_0\omega\boldsymbol{k}\times\boldsymbol{B}(\boldsymbol{k},\omega) + \boldsymbol{k}\times[\boldsymbol{k}\times\boldsymbol{P}(\boldsymbol{k},\omega)]$$

$$\rightarrow [\overbrace{\boldsymbol{k}\cdot\boldsymbol{D}(\boldsymbol{k},\omega)}^{0}]\boldsymbol{k} - k^2\boldsymbol{D}(\boldsymbol{k},\omega) = -\mu_0\varepsilon_0\omega^2\boldsymbol{D}(\boldsymbol{k},\omega) + \varepsilon_0\omega\boldsymbol{k}\times\boldsymbol{M}(\boldsymbol{k},\omega) + \boldsymbol{k}\times[\boldsymbol{k}\times\boldsymbol{P}(\boldsymbol{k},\omega)]$$

$$\rightarrow \quad (k^2 - \omega^2/c^2)\boldsymbol{D}(\boldsymbol{k},\omega) = -\varepsilon_0\omega\boldsymbol{k}\times\boldsymbol{M}(\boldsymbol{k},\omega) - \boldsymbol{k}\times[\boldsymbol{k}\times\boldsymbol{P}(\boldsymbol{k},\omega)]$$

$$\rightarrow \quad \boldsymbol{D}(\boldsymbol{k},\omega) = -\varepsilon_0\omega\boldsymbol{k}\times\frac{\boldsymbol{M}(\boldsymbol{k},\omega)}{k^2-(\omega/c)^2} - \boldsymbol{k}\times\left[\boldsymbol{k}\times\frac{\boldsymbol{P}(\boldsymbol{k},\omega)}{k^2-(\omega/c)^2}\right]. \tag{B2}$$

Equations (B1) and (B2) are used in Sec.8 to define the Hertz vector potentials.

## Appendix C

This appendix provides a listing of some important properties of the Hermite equation and its corresponding polynomial solutions. The source is Gradshteyn and Ryzhik's *Table of Integrals, Series, and Products*.[26]

Hermite's equation: $$f''(x) - 2xf'(x) + 2nf(x) = 0. \quad \text{(G\&R 8.959-1)}$$

Hermite Polynomials: $$H_n(x) = (-1)^n\exp(x^2)\frac{\mathrm{d}^n}{\mathrm{d}x^n}\exp(-x^2). \quad \text{(G\&R 8.950-1)}$$

Normalization: $$\int_{-\infty}^{\infty}\exp(-x^2)\,H_m(x)H_n(x)\mathrm{d}x = \begin{cases}0 & (m\neq n),\\ \sqrt{\pi}\,n!\,2^n & (m=n).\end{cases} \quad \text{(G\&R 7.374-1)}$$

Fourier transformation: $$\int_{-\infty}^{\infty}H_n(x)e^{-\frac{1}{2}x^2}e^{\pm\mathrm{i}sx}\mathrm{d}x = (\pm\mathrm{i})^n\sqrt{2\pi}H_n(s)e^{-\frac{1}{2}s^2}. \quad \text{(G\&R 7.376-1)}$$

Methods of solving Hermite's equation as well as proofs of various identities can be found in most books on *Mathematical Methods in Science and Engineering*; see, for example, [25].

## Appendix D

The asymptotic behavior of the inverse Fourier transform of $F(s) = \exp[2s_0/(s^2 - s_0^2)]$, defined over $|s| \le s_0$, is determined by recognizing that, when $x \to \infty$, the Fourier integral needs

to be evaluated only in the vicinity of the end-points $s = \pm s_0$, where the following approximations are admissible:

$$\begin{aligned}
f(x) &= \int_{-s_0}^{s_0} e^{-1/(s_0-s)} e^{-1/(s_0+s)} e^{\mathrm{i}2\pi xs} \mathrm{d}s \\
&\cong e^{-1/(2s_0)} \int_{-s_0}^{\infty} e^{-1/(s_0+s)} e^{\mathrm{i}2\pi xs} \mathrm{d}s + e^{-1/(2s_0)} \int_{-\infty}^{s_0} e^{-1/(s_0-s)} e^{\mathrm{i}2\pi xs} \mathrm{d}s \\
&= e^{-1/(2s_0)} \left[ \int_0^{\infty} e^{-1/s} e^{\mathrm{i}2\pi x(s-s_0)} \mathrm{d}s + \int_0^{\infty} e^{-1/s} e^{\mathrm{i}2\pi x(s_0-s)} \mathrm{d}s \right] \\
&= 2e^{-1/(2s_0)} \mathrm{Re} \left[ e^{\mathrm{i}2\pi s_0 x} \int_0^{\infty} e^{-(1/s)-\mathrm{i}2\pi xs} \mathrm{d}s \right] \\
&= 4e^{-1/(2s_0)} \mathrm{Re} \left[ \frac{e^{\mathrm{i}2\pi s_0 x}}{\sqrt{\mathrm{i}2\pi x}} K_1\left(2\sqrt{\mathrm{i}2\pi x}\right) \right] \\
&\cong 4e^{-1/(2s_0)} \mathrm{Re} \left[ \frac{e^{\mathrm{i}2\pi s_0 x}}{\sqrt{\mathrm{i}2\pi x}} \times \frac{\sqrt{\pi}}{2\sqrt[4]{\mathrm{i}2\pi x}} e^{-2\sqrt{\mathrm{i}2\pi x}} \right] \\
&= \left(\frac{2}{\pi x^3}\right)^{1/4} \exp\left(-\sqrt{4\pi x} - \frac{1}{2s_0}\right) \cos\left(\sqrt{4\pi x} - 2\pi s_0 x + \frac{3\pi}{8}\right). \qquad \text{(D1)}
\end{aligned}$$

The integral appearing on the 4th line of the above equation is listed in Gradshteyn and Ryzhik's Table of Integrals as equation $(3.324-1)$.[26] The Bessel function asymptotics on the 6th line, namely, $K_1(z) \sim \sqrt{\pi/(2z)}\, e^{-z}$, appear in the same handbook as equation $(8.451-6)$.

## Appendix E

The half-derivative of the rectangular pulse function $\mathrm{rect}(x)$ is determined as follows:

$$\begin{aligned}
\frac{\mathrm{d}^{1/2}}{\mathrm{d}x^{1/2}} \mathrm{rect}(x) &= \int_{-\infty}^{\infty} (\mathrm{i}2\pi s)^{1/2} \,\mathrm{sinc}(s) \exp(\mathrm{i}2\pi sx)\, \mathrm{d}s = \int_{-\infty}^{\infty} \frac{\exp[\mathrm{i}2\pi s(x+1/2)] - \exp[\mathrm{i}2\pi s(x-1/2)]}{(\mathrm{i}2\pi s)^{1/2}} \mathrm{d}s \\
&= \sqrt{2/\pi} \exp(\mathrm{i}\pi/4) \int_0^{\infty} \{\exp[-\mathrm{i}2\pi(x+1/2)\zeta^2] - \exp[-\mathrm{i}2\pi(x-1/2)\zeta^2]\} \mathrm{d}\zeta \\
&\quad + \sqrt{2/\pi} \exp(-\mathrm{i}\pi/4) \int_0^{\infty} \{\exp[\mathrm{i}2\pi(x+1/2)\zeta^2] - \exp[\mathrm{i}2\pi(x-1/2)\zeta^2]\} \mathrm{d}\zeta \\
&= \sqrt{2/\pi} \begin{cases} 0; & (x < -1/2) \\ \int_0^{\infty} [\exp(-2\pi|x+1/2|r^2) - \mathrm{i}\exp(-2\pi|x-1/2|r^2)] \mathrm{d}r & \\ \quad + \int_0^{\infty} [\exp(-2\pi|x+1/2|r^2) + \mathrm{i}\exp(-2\pi|x-1/2|r^2)] \mathrm{d}r; & (|x| < 1/2) \\ 2\int_0^{\infty} \{\exp[-2\pi(x+1/2)r^2] - \exp[-2\pi(x-1/2)r^2]\} \mathrm{d}r; & (x > 1/2) \end{cases} \\
&= \begin{cases} 0; & (x < -1/2), \\ \frac{1}{\sqrt{\pi(x+1/2)}}; & (|x| < 1/2), \\ \frac{1}{\sqrt{\pi(x+1/2)}} - \frac{1}{\sqrt{\pi(x-1/2)}}; & (x > 1/2) \end{cases} = \frac{\mathrm{step}(x+1/2)}{\sqrt{\pi(x+1/2)}} - \frac{\mathrm{step}(x-1/2)}{\sqrt{\pi(x-1/2)}}. \qquad \text{(E1)}
\end{aligned}$$

In the first step of this derivation, we have changed the variable to $\zeta = \sqrt{-s}$ for negative $s$, and to $\zeta = \sqrt{s}$ for positive $s$. In the second step, the integrals are evaluated on $\pm 45°$ lines in the complex plane, where $\zeta = re^{\pm \mathrm{i}\pi/4}$.